\documentclass[journal,12pt,onecolumn, draftclsnofoot]{IEEEtran}
\usepackage[pdftex]{graphicx}

\usepackage{amsmath}
\usepackage{amsthm}
\usepackage[flushleft]{threeparttable}
\usepackage{array}
\usepackage{booktabs}
\usepackage{bbm}
\newcommand{\PreserveBackslash}[1]{\let\temp=\\#1\let\\=\temp}
\newcolumntype{C}[1]{>{\PreserveBackslash\centering}p{#1}}
\newcolumntype{R}[1]{>{\PreserveBackslash\raggedleft}p{#1}}
\newcolumntype{L}[1]{>{\PreserveBackslash\raggedright}p{#1}}

\usepackage{bm}
\usepackage{graphicx,subfigure}
\usepackage{algorithm}
\usepackage{cite}
\usepackage{amsmath}
\usepackage{amssymb}
\usepackage{psfrag}
\usepackage{algpseudocode}
\usepackage{empheq}
\usepackage{latexsym, amsmath, subfigure, color, amsfonts, amssymb,graphicx}
\usepackage{wrapfig}
\usepackage{enumitem}
\usepackage{cleveref}
\newtheorem{Theorem}{Theorem}
\newtheorem{Lemma}{Lemma}
\newtheorem{Definition}{Definition}

\newtheorem{Remark}{Remark}
\IEEEoverridecommandlockouts
\usepackage{tabularx}
\DeclareMathOperator*{\argmin}{\arg\!\min}
\DeclareMathOperator*{\argmax}{\arg\!\max}

\usepackage[T1]{fontenc}
\usepackage{etoolbox}

\makeatletter
\def\hlinewd#1{%
\noalign{\ifnum0=`}\fi\hrule \@height #1 %
\futurelet\reserved@a\@xhline}
\patchcmd{\maketitle}{\@fnsymbol}{\@alph}{}{}  % Footnote numbers from symbols to small letters
\makeatother

\title{Audience-Retention-Rate-Aware \\ Caching and Coded Video Delivery with \\ Asynchronous Demands}
\author{
Qianqian Yang, Mohammad Mohammadi Amiri, and\thanks{The authors are with Imperial College London, London SW7 2AZ, U.K. (e-mail: q.yang14@imperial.ac.uk; m.mohammadi-amiri15@imperial.ac.uk; d.gunduz@imperial.ac.uk).

This paper was presented in part at the IEEE International Conference on Communications, Workshop on Advanced Caching for Wireless Networks, Paris, France, May 2017 \cite{yangaudience}. }
  \and
  Deniz G\"und\"uz
}

\date{}
\begin{document}

\maketitle

\begin{abstract}
Most\makeatletter{\renewcommand*{\@makefnmark}{}\footnotetext{This work received support from EC H2020-MSCA-ITN-2015 project SCAVENGE under grant number 675891, and from the European Research Council project BEACON under grant number 677854.}\makeatother} results on coded caching focus on a static scenario, in which a fixed number of users synchronously place their requests from a content library, and the performance is measured in terms of the latency in satisfying all of these demands. In practice, however, users start watching an online video content asynchronously over time, and often abort watching a video before it is completed. The latter behaviour is captured by the notion of \textit{audience retention rate}, which measures the portion of a video content watched on average. In order to bring coded caching one step closer to practice, asynchronous user demands are considered in this paper, by allowing user demands to arrive randomly over time, and both the popularity of video files, and the audience retention rates are taken into account. A decentralized \textit{partial coded caching} (PCC) scheme is proposed, together with two cache allocation schemes; namely the \textit{optimal cache allocation} (OCA) and the \textit{popularity-based cache allocation} (PCA), which allocate users' caches among different chunks of the video files in the library. Numerical results validate that the proposed PCC scheme, either with OCA or PCA, outperforms conventional uncoded caching as well as the state-of-the-art decentralized caching schemes, which consider only the file popularities, and are designed for synchronous demand arrivals. An information-theoretical lower bound on the average delivery rate is also presented. 
\end{abstract}

\section{Introduction}\label{intro}
The ever-increasing demand for video services has been the main driver for the recent explosive growth of wireless data traffic. A key feature of video services is that a small portion of highly popular contents dominate the traffic~\cite{zeni2013youstatanalyzer}. This led to the idea of prefetching popular contents over off-peak traffic periods, or at better channel conditions, and storing them at the network edge~\cite{GolrezaeiFemtocaching}, or even directly at user devices~\cite{GregoryDtoD, MaddahAliCentralized}, referred to as \textit{proactive caching}. Proactive caching can alleviate both the growing traffic load on the backhaul links and the associated latency; and it becomes more viable thanks to the decreasing cost of memory (see \cite{GolrezaeiFemtocaching, GregoryDtoD, MaddahAliCentralized,samuel2017}, and references therein).

In proactive caching, during off-peak traffic periods, users' caches are filled, without the knowledge of future user demands, referred to as the \textit{placement phase}. Users' demands are revealed during the peak traffic period, and are satisfied simultaneously over the \textit{delivery phase}. Traditional uncoded caching schemes adopt orthogonal unicast transmissions, and the caching gain is limited by the capacity of each user's local cache memory. On the other hand, \textit{coded caching}, a novel caching paradigm introduced in \cite{MaddahAliCentralized}, exploits the cache resources across the network by jointly optimizing the two phases in order to create and exploit coded multicasting opportunities, even among distinct user requests. It is shown in \cite{MaddahAliCentralized} that coded caching provides a global caching gain, which depends on the total cache capacity in the network. Coded caching and delivery has ignited intense research activities in recent years \cite{MaddahAliDecentralized,MohammadQianDenizITW,MohammadDenizTCom,JiArXivNonuniform,NiesenNonuniform,PedarsaniOnlineCaching,yang2016coded, DistinctAmiriYangGunduz}.

There are two limitations of the current literature on coded caching that we address in this paper: The first is the assumption that all the users in the system request their desired files simultaneously at the beginning of the delivery phase\footnote{A slotted time model is considered in \cite{PedarsaniOnlineCaching} to take the dynamics of library into account. However, users are still assumed to place their requests simultaneously at each time slot.}. Moreover, it is assumed that the users request entire files\footnote{It is assumed in \cite{yang2016coded} that users may request the same file at different resolutions. However, this model is still limited to considering a complete request from the library for each user.}. However, in practice, users rarely request and watch an entire video content, and different user requests may arrive at different time instants, and each user may abort watching a certain video content after a random duration. A recent report \cite{zeni2013youstatanalyzer} suggests that, users on average watch $60\%$ of their requested files from a trace of $7000$ Youtube videos, and the number of views varies over different videos as well as different parts of each video. This phenomena is captured by the notion of \textit{audience retention rate}, introduced by the mainstream online video platforms, such as Youtube and Netflix, to model the popularities of different parts of available content, and it is provided to content generators to better understand user engagement with generated content. For efficient caching and delivery, this nonuniform viewing behaviour calls for \textit{partial caching}, where only the most viewed portion of each video file is cached. Audience retention rate aware partial caching is shown to improve the performance of uncoded caching in \cite{maggi2015adapting}. 

Here, we investigate coded caching of video files taking into account the audience retention rate for each video. We consider that each video file consists of equal-length chunks, and the audience retention rate of each chunk is the fraction of users watching this chunk among total views of the corresponding video. Also, in contrast to the literature on coded caching, where users are assumed to reveal their demands simultaneously, we consider a more realistic dynamic demand arrival model, where users randomly join the delivery over time, and leave after watching a random number of video chunks. Taking both the asynchronous demand arrivals and the audience retention rate into account, we propose a novel decentralized caching scheme, referred to as \textit{partial coded caching} (PCC). We derive a closed-form expression for the achievable average delivery rate over all possible demand combinations. Two different cache allocation schemes are proposed to allocate users' caches to different chunks, namely \textit{optimal cache allocation} (OCA) and \textit{popularity based cache allocation} (PCA). We also derive an information-theoretic lower bound on the achievable average delivery rate. Note that the coded caching problem with different file popularities, studied in \cite{JiArXivNonuniform, emreicc2018}, is a special case of the problem considered in this paper obtained by setting the audience retention rates of all the chunks to one, and assuming all the demands arrive simultaneously. Numerical results indicate that the proposed audience retention rate aware partial coded caching scheme achieves a better delivery rate than both uncoded caching and the scheme proposed in \cite{JiArXivNonuniform} adapted to the current setting.

The rest of this paper is organized as follows. The system model is introduced in Section \ref{system}. In Section \ref{SecPCCScheme}, we introduce the proposed partial coded caching scheme, and analyze its performance in terms of the average delivery rate. We present a lower bound on the performance of the system in Section \ref{SecLowerBound}. Numerical results are presented in Section \ref{section:numerical}. Finally, we conclude the paper in Section \ref{SecConc}, followed by the Appendices with the complete proofs.

\textit{Notations:} We denote the set of $t$-bit binary sequences by $[2^t]$, and the set of all binary sequences by $[2^*]$. The set of integers $\left\{ i, ..., j \right\}$, where $i \le j$, is denoted by $\left[ i:j \right]$, while, $\left\{1, ..., j \right\}$ is denoted by $\left[j \right]$. For sets $\mathcal{A}$ and $\mathcal{B}$, we define $\mathcal{A} \backslash \mathcal{B}\triangleq\{x: x \in \mathcal{A}, x\notin \mathcal{B}\}$, and $\left| \mathcal{A} \right|$ denotes the cardinality of $\mathcal{A}$. Notation $\overline \oplus$ represents the bitwise XOR operation, where the arguments are zero-padded to have equal length. For two positive integers $i,j$, $i \le j$, $K_{i:j}$ denotes $(K_i, ..., K_j)$; while $K_{[j]}$ denotes $(K_1, ..., K_j)$. For event $E$, $\mathbbm{1}\{E\}=1$ if $E$ is true; and $\mathbbm{1}\{E\}=0$, otherwise. $\binom{j}{i}$ represents the binomial coefficient if $j\geq i$; and $\binom{j}{i}=0$, otherwise. $\mathbb{R}$ and $\mathbb{N}$ denote the sets of real numbers and positive integers, respectively.

\section{System Model}\label{system}
We consider a server holding a library of $N$ popular video files, denoted by $\mathcal{F}=\{W_1, ..., W_N\}$. We assume, for simplicity, that all the files have the same size of $F$ bits. Each file consists of $B$ chunks of equal size, i.e., $F/B$ bits each, which is determined by various factors in practical applications, such as the frame size, display settings of user devices,  etc.~\cite{wang2015optimal}. We denote by $W_{ij}$ the $j$th chunk of file $W_i$. 

In the placement phase, each user pre-fetches data from the server to fill its cache of size $MF$ bits. We consider a dynamic delivery phase; that is, users arrive randomly, request a random video from the library, watch a random number of chunks of that video, and leave the system. Active users at any time instant are connected to the server through an error-free shared link.  

We consider a slotted time model, where the beginning of the delivery phase is marked as $t=0$, and the unit time interval $(t-1, t]$ is referred to as time slot $t$, $t\in \mathbb{N}$. We assume that a user consumes exactly one chunk of a video file in one time slot. We denote the number of new demands that arrive during time slot $t$ as $a_t$, where $a_t$ is independently and identically distributed (i.i.d.) according to $P_A$ over set $\mathcal{A}$, which is bounded, i.e., $\mathcal{A}=\{0, 1, ..., A_{\mathrm{max}}\}$, since only a limited number of new users can be admitted at each time slot. Each demand corresponds to a file from $\mathcal{F}$ which is i.i.d. according the \textit{popularity distribution} $\mathbf{p}\triangleq (p_1, ..., p_N)$.  

Unlike the current literature on coded caching, we do not necessarily deliver the requested contents in their entirety, as users often quit watching a video file before completion. Therefore, in our model, users are initially delivered only the first chunks of their desired video files. Their demands of subsequent chunks are only revealed after receiving the previous ones, unless they abort watching the video. Specifically, the first chunks of the $a_t$ demands that have arrived in slot $t$ are delivered during slot $t+1$, and then the corresponding $a_t$ users decide to continue watching or not after having received the first chunks. Those who have decided to continue watching are served the second chunks of their requested files during slot $t+2$. In the same manner, having received $j$th chunks during slot $t+j$, the users who continue watching are delivered the $(j+1)$th chunks during slot $t+j+1$, for $j \in [B-1]$. We note that the first chunks of the requested files are always delivered. We also note that, according to this model, at any time slot $t$, the server may be serving demands that have arrived at time slots $t-B, t-B+1, \ldots, t-1$.

\begin{figure}[!t]
\centering
\includegraphics[scale=0.88]{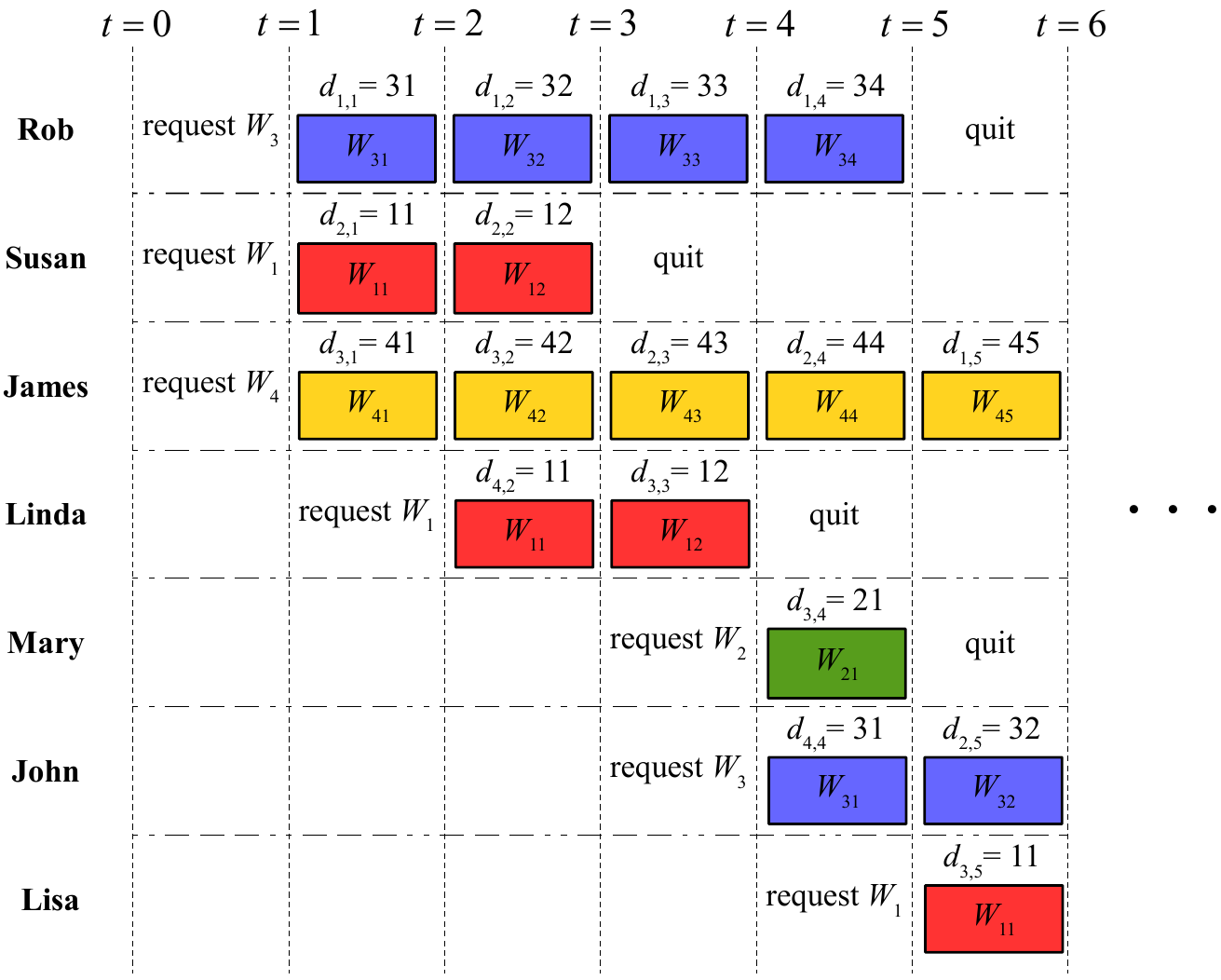}
\caption{Illustration of the demand arrivals for an asynchronous caching system with $N \ge 4$ files and $A_{\rm{max}} \ge 3$ for time slots $t=1$ to $6$ of the delivery phase. In the caching setting under consideration, we have $a_1=3$, $a_2=1$, $a_3=0$, $a_4=2$, $a_5=1$ demands, and $K^{(1)} = 3$, $K^{(2)} = 4$, $K^{(3)} = 3$, $K^{(4)} = 4$, $K^{(5)} = 3$ users served at each time slot.}
\label{SystemModel}
\end{figure}

To model this, we employ the notion of \textit{audience retention rate}, defined as the fraction of users that request chunk $W_{ij}$ among all the users that have requested $W_i$, denoted by $p_{ij}$, for $i \in [N]$ and $j \in [B]$~\cite{maggi2015adapting}. Alternatively, we can regard $p_{ij}$ as the probability that a user who requested video $W_i$ will watch the $j$th chunk\footnote{Here we assume that a user cannot skip chunk $W_{ij}$ for some $j \in [B]$, and request a later chunk $k$, for $k>j$. Once a user does not request chunk $W_{ij}$, it leaves the system, and does not receive any further content.}. Accordingly, $p_{ij}$ is non-increasing in $j$, i.e., $1=p_{i1}\geq p_{i2}\geq \cdots \geq p_{iB}$, which characterizes a realistic viewing model that users start watching videos from the beginning and abort after watching a random number of chunks in order. We let $\mathbf{P}=\{p_{ij}, i \in [N], j \in [B]\}$ denote the \textit{retention rate matrix} for all the chunks in the library, which is time-invariant and identical for all the users. We refer to $p_ip_{ij}$ as the popularity of chunk $W_{ij}$ in the sense that it denotes the probability that chunk $W_{ij}$ will be requested by a user that joins the system.

During the placement phase, each user fills its cache as an arbitrary function of the library $\mathcal{F}$, the file popularity vector $\mathbf{p}$, and the retention rate matrix $\mathbf{P}$, subject to its cache capacity of $MF$ bits. We emphasize that the knowledge of the future requests is not available during the placement phase, and only during the placement phase the contents in the caches are updated. We also note that the placement phase is performed in a decentralized manner; that is, coordination among the users during this phase is not possible since the server does not know when a user is going to make a request in advance.

The delivery phase begins once the users start requesting files, and as described above, is performed over many time slots. At each slot $t$, the server serves all the active users in the system, those from slot $t-1$ that continue watching their requested contents, as well as the new arrivals. Denote by $K^{(t)}$ the total number of active users to be served at time slot $t$, and $K^{(t)}_j$ the number of users among the $K^{(t)}$ active users requesting their $j$th chunks. All the active users are re-indexed at the beginning of slot $t$ as $[K^{(t)}]$ in a way that users $\sum_{h=1}^{j-1} K^{(t)}_h+1$ to $\sum_{h=1}^{j} K^{(t)}_h$ are requesting their $j$th chunks, $j=1, ..., B$. The cache content of the $k$th user, for $k \in [K^{(t)}]$, is denoted by $Z_k^{(t)}$. Let $d_{k, t}$ denote the index of the chunk requested by user $k$, which needs to be delivered at slot $t$, $d_{k, t} \in \{ij: i\in [N], j \in [B]\}$. We remark that if user $k$ has joined the system at slot $t'$, then $d_{k, t} \in \{ij: i\in [N], j=t-t'\}$, i.e., at slot $t$, the $(t-t')$th chunk of her request will be delivered to user $k$. Let $\mathbf{d}_t\triangleq(d_{1, t}, ..., d_{K^{(t)}, t})$ denote the demand vector at slot $t$, and $\mathcal{D}_t\triangleq\{W_{d_{1, t}}, ..., W_{d_{K^{(t)}, t}}\}$ denote the set of requested chunks. Then, to satisfy all these requests at slot $t$, the server sends a message of length $R_{\mathbf{d}_t}F/B$ bits over the shared link, which is a function of the library $\mathcal{F}$, the demand vector $\mathbf{d}_t$, and the cache contents of the active users $Z^{(t)}_1$, ..., $Z^{(t)}_{K^{(t)}}$. User $k$ recovers chunk $W_{d_{k, t}}$ at the end of slot $t$ from the transmitted message and the contents in her local cache. We are interested in long-term the average delivery rate 
\begin{equation}\label{defi1}
R\triangleq \lim\limits_{T\rightarrow \infty}\mathbb{E} \left[\frac{1}{T}\sum\limits_{t=1}^TR_{\mathbf{d}_t}\right],~~ \mathrm{for}~t\geq B,
\end{equation}
where the expectation is taken over all demand realizations $\mathbf{d}_t$ distributed according to $P_A$, $\mathbf{p}$, and $\mathbf{P}$. 

It is easy to see that $\mathbf{d}_t$ is a Markov chain, and since the cached contents of active users are constant throughout the delivery phase, $R_{\mathbf{d}_t}$ at time $t$ depends only on the current state $\mathbf{d}_t$; therefore, the long-term average rate in \eqref{defi1} can be obtained by simply evaluating $\mathbb{E} [\sum\nolimits_{t=1}^TR_{\mathbf{d}_t}]$ in the steady state demand distribution, which depends on $P_A$, $\mathbf{p}$, and $\mathbf{P}$.
\begin{Definition}
A cache capacity-average rate pair $(M,R)$ is \textit{achievable} for the caching system described above, if there exists a caching and delivery scheme with cache capacity $M$ at each user and average rate ${R}$ such that for any demand realization $\mathbf{d}_t$, $\forall t$,
\begin{equation}
\lim_{F/B \rightarrow \infty} \Pr \left\{\bigcup\nolimits_{k} \Big\{\hat{W}_{d_{t, k}}\neq W_{d_{t, k}}\Big\}\right\}=0,
\end{equation} 
where $\hat{W}_{d_{k,t}}$ denotes the reconstruction of $W_{d_{k,t}}$ at user $k$ at the end of time slot $t$.
\end{Definition}
We define $R^*(M) \buildrel \Delta \over = \min \left\{ \mbox{$R:\left( M,R \right)$ is achievable} \right\}$ to express the tradeoff between the cache capacity and the average delivery rate. The goal in this paper is to characterize this trade-off. 

\begin{Remark}
The delivery rate $R_{\mathbf{d}_t}$ as defined above (following \cite{MaddahAliDecentralized}) refers to the total number of bits that must be delivered in order to satisfy all user demands in time slot $t$. Therefore, it can be considered as a measure of latency, rather than the more classical communication rate concept. In our setting, however, we consider a slotted system; hence, the duration of each time slot is considered fixed according to the display duration of one chunk of a video file. Accordingly, $R_{\mathbf{d}_t}$ can be considered as a measure of the bandwidth/capacity required to satisfy all user demands within a time slot duration to guarantee the streaming of video files without stalling. 
\end{Remark}

\section{Partial Coded Caching (PCC)}\label{SecPCCScheme} 
Here we first present our coded caching scheme, referred to as \textit{partial coded caching} (PCC), along with an example, and then derive its average delivery rate. We remark that the number of bits and the delivery rate mentioned in the sequel are both normalized by $F/B$.

\subsection{Placement Phase}\label{placement}
During the placement phase, each user selects an independent random subset of $q_{ij}F/B$ bits from $W_{ij}$ to fill its cache, where $0 \leq q_{ij}\leq 1$, such that $\sum\nolimits_{i=1}^N\sum\nolimits_{j=1}^Bq_{ij}=MB$, which, for large $F$, satisfies the cache capacity constraint with high probability. We refer to $\mathbf{Q}=\{q_{ij}, i\in [N], j \in [B]\}$ as the \textit{cache content distribution}, which will be optimized in order to minimize the average delivery rate. The optimization of $\mathbf{Q}$ is studied in Section~\ref{cache}. 
 
\subsection{Delivery Phase}\label{delivery}
\begin{algorithm}[t]
\caption{Random Delivery}
\label{randomdelivery}
\begin{algorithmic}[1]
\For{$W_{ij}\in \mathcal{D}_t$}
\State {Server sends enough random linear combinations of the bits of file $W_{ij}$ to enable the users demanding it to decode it.}
\EndFor
\end{algorithmic}
\end{algorithm}
\begin{algorithm}[t]
\caption{Delivery scheme during time slot $t$ based on \cite[Algorithm 1]{MaddahAliDecentralized}}
\label{deliveryscheme1}
\begin{algorithmic}[1]
\State{Delivering the missing bits that are in the cache of any subset of users in $\mathcal{K}_t$:}
\For{$z=0, ..., K^{(t)}$}
\For{$\mathcal{P} \subset  [K^{(t)}]$: $|\mathcal{P}|=z$}
\State{Send ${\overline  \bigoplus}_{k\in \mathcal{P}}W^t_{d_{k, t}, \mathcal{P}\setminus\{k\}}$}
\EndFor
\EndFor
\end{algorithmic}
\end{algorithm}
\begin{algorithm}[!ht]
\caption{Delivery scheme during time slot $t$}
\label{deliveryscheme}
\begin{algorithmic}[1]
\State{\textbf{PART 1}: Delivering the missing bits that are not in the cache of any user in $[K^{(t)}]$:}
\For {$W_{ij}\in \mathcal{D}_t$}
\State{Send $W^t_{ij, \emptyset}$}
\EndFor
%\Statex
\State{\textbf{PART 2}: Delivering the missing bits that are in the cache of only one user in $[K^{(t)}]$; the one among \textbf{PART 2.1} and \textbf{PART 2.2} that requires a smaller delivery rate is executed:}
\State{\textbf{PART 2.1:}}
\For{$\mathcal{P} \subset [K^{(t)}]$: $|\mathcal{P}|=2$}
\State{Send ${\overline  \bigoplus}_{k\in \mathcal{P}}W^t_{d_{k, t}, \mathcal{P}\setminus\{k\}}$}
\EndFor

\State{\textbf{PART 2.2:}}
\For{$W_{ij}\in \mathcal{D}_t$}
\State{Send $\bigcup\nolimits_{k =1}^{K^{(t)}-1}W^t_{ij, \{k\}}\overline{\oplus}W^t_{i j,\{k+1\}}$}
\EndFor
%\Statex
\State{\textbf{PART 3}: Delivering the missing bits that are in the cache of more than one user in $[K^{(t)}]$:}
\For{$\mathcal{P} \subset  [K^{(t)}]$: $|\mathcal{P}|>2$}
\State{Send ${\overline  \bigoplus}_{k\in \mathcal{P}}W^t_{d_{k, t}, \mathcal{P}\setminus\{k\}}$}
\EndFor
\end{algorithmic}
\end{algorithm}
As described in the system model, the delivery phase is performed over different time slots, according to the current demand configuration specified by $\mathbf{d}_t$ during each time slot $t$, and cache contents $Z^{(t)}_1, ..., Z^{(t)}_{K^{(t)}}$, where $K^{(t)}$ denotes the number of active users at slot $t$. We emphasize that users' requests for the $j$-th chunks are revealed only after they receive the first $j-1$ chunks, and in the delivery phase a user is not served a chunk before requesting it. For $\mathcal{S} \subset [K^{(t)}]$, we denote by $W^t_{ij, \mathcal{S}}$ the bits of chunk $W_{ij}$ that are exclusively cached by the users in $\mathcal{S}$ (i.e., not cached by any of the users in $[K^{(t)}]\setminus \mathcal{S}$). We note that $W^t_{ij, \mathcal{S}}\not\equiv W^{t'}_{ij, \mathcal{S}}$ for $t\neq t'$, $t, t' \in \mathbbm{N}$, since a different set of users may be active at each time slot; and thus, $\mathcal{S}$ may refer to a different subset of users at different time slots. 

In Algorithm \ref{randomdelivery}, we present the \textit{Random Delivery} (RAN) scheme, which simply delivers random linear combinations of the bits of a chunk until it is decoded by the requesting user. This scheme has been considered in \cite{MaddahAliDecentralized, DistinctAmiriYangGunduz, NiesenNonuniform} as an alternative delivery procedure although it is known to perform poorly in general compared to coded delivery. 

The second delivery scheme considered here is presented in Algorithm \ref{deliveryscheme1}. We will refer to it as the MAN scheme as it is based on \cite[Algorithm 1]{MaddahAliDecentralized}. We remark that, here we use operation ${\overline  \bigoplus}$ instead of the $\bigoplus$ in \cite[Algorithm 1]{MaddahAliDecentralized}. Note also that, although we include $z=0$ in Algorithm \ref{deliveryscheme1} compared to \cite[Algorithm 1]{MaddahAliDecentralized}, it does not increase the delivery rate since when $\mathcal{P}$ is an empty set, no bit is sent by the server according to Algorithm \ref{deliveryscheme1}. 

The novel PCC scheme is presented in Algorithm~\ref{deliveryscheme}. PCC was first introduced in~\cite{yangaudience} for synchronous user demands. Here we will optimize and analyze its performance for asynchronous user demands.

\begin{Remark}
The coded delivery scheme in \cite[Algorithm 1]{DistinctAmiriYangGunduz} in general can achieve a lower delivery rate than the above schemes for the same demand combination. However, the average delivery rate of the scheme in \cite[Algorithm 1]{DistinctAmiriYangGunduz} does not lend itself to a closed-form expression; and therefore, it would not be possible to optimize the cache allocation to minimize the average delivery rate. The proposed PCC scheme, on the other hand, allows the optimization of cache allocation functions, and outperforms the state-of-the-art results for coded caching with non-uniform file popularities, as it will be shown in the sequel.
\end{Remark}

%Before providing a detailed analysis of the average delivery rate achieved by the above schemes, we will explain them on an example.

\subsection{Example}\label{sec:example}
Here we explain the coded delivery schemes outlined in Algorithms \ref{deliveryscheme1} and \ref{deliveryscheme} for an arbitrary time slot $t \ge 2$, assuming $N = 3$ files in the library, each consisting of $B = 2$ chunks. Every user in the system performs decentralized placement as described in Section \ref{placement}. Assume that $K^{(t)} = 5$ users are active in slot $t$: $3$ new users (users 1, 2 and 3) will be served with the first chunks of their demands, while $2$ of the users (users 4 and 5) that started watching their demands in the previous time slot continue watching their respective video files; and thus, will be served the second chunks of their demands. Assume that users 1, 2 and 3 request files $W_2$, $W_1$ and $W_2$, respectively, while users 4 and 5 have requested $W_3$ and $W_2$, respectively. Thus, the demand vector is $\mathbf{d}_t = \{ 21,11,21,32,22 \}$, and $\mathcal{D}_t = \left\{ W_{21},W_{11},W_{32},W_{22} \right\}$, where, for $W_{ij} \in \mathcal{D}_t$, we have $W_{ij} = \bigcup\nolimits_{\mathcal{S}\subset [5]} W^t_{ij,\mathcal{S}}$. We note that, given the placement phase presented in Section \ref{placement}, for $F$ large enough, and $\mathcal{S}\subset [5]$, the size of $W^t_{ij,\mathcal{S}}$, is given by $\left( q_{ij} \right)^{\left| \mathcal{S} \right|} \left( 1-q_{ij} \right)^{5-\left| \mathcal{S} \right|}F/2$ bits. Also note that, in order to obtain the intended chunk, user $k \in [5]$ should recover $\bigcup\nolimits_{\mathcal{S}\subset [5]: k \notin \mathcal{S}} W^t_{d_{k,t},\mathcal{S}}$.

We first consider the delivery scheme of Algorithm \ref{deliveryscheme1}. The following contents are delivered for $z=1, ..., 5$: 
\begin{itemize}
\item For $z=1$:
\end{itemize}
\begin{align}\label{DeliveredContentsAlg1z1Example}
\footnotesize
W^t_{21,\emptyset},W^t_{11,\emptyset},W^t_{21,\emptyset},W^t_{32,\emptyset},W^t_{22,\emptyset}. 
\normalsize
\end{align}
\begin{itemize}
\item For $z=2$:
\end{itemize}
\begin{align}\label{DeliveredContentsAlg1z2Example}
\footnotesize
& W^t_{21,\{ 2 \}} {\overline  \oplus} W^t_{11,\{ 1 \}}, W^t_{21,\{ 3 \}} {\overline  \oplus} W^t_{21,\{ 1 \}}, W^t_{21,\{ 4 \}} {\overline  \oplus} W^t_{32,\{ 1 \}}, W^t_{21,\{ 5 \}} {\overline  \oplus} W^t_{22,\{ 1 \}}, W^t_{11,\{ 3 \}} {\overline  \oplus} W^t_{21,\{ 2 \}},\nonumber\\
& W^t_{11,\{ 4 \}} {\overline  \oplus} W^t_{32,\{ 2 \}},W^t_{11,\{ 5 \}} {\overline  \oplus} W^t_{22,\{ 2 \}}, W^t_{21,\{ 4 \}} {\overline  \oplus} W^t_{32,\{ 3 \}},W^t_{21,\{ 5 \}} {\overline  \oplus} W^t_{22,\{ 3 \}},W^t_{32,\{ 5 \}} {\overline  \oplus} W^t_{22,\{ 4 \}}.
\normalsize
\end{align}
\begin{itemize}
\item For $z=3$:
\end{itemize}
\begin{align}\label{DeliveredContentsAlg1z3Example}
\footnotesize
& W^t_{21,\{ 2,3 \}} {\overline  \oplus} W^t_{11,\{ 1,3 \}} {\overline  \oplus} W^t_{21,\{ 1,2 \}}, W^t_{21,\{ 2,4 \}} {\overline  \oplus} W^t_{11,\{ 1,4 \}} {\overline  \oplus} W^t_{32,\{ 1,2 \}}, W^t_{21,\{ 2,5 \}} {\overline  \oplus} W^t_{11,\{ 1,5 \}} {\overline  \oplus} W^t_{22,\{ 1,2 \}},\nonumber\\
&  W^t_{21,\{ 3,4 \}} {\overline  \oplus} W^t_{21,\{ 1,4 \}} {\overline  \oplus} W^t_{32,\{ 1,3 \}}, W^t_{21,\{ 3,5 \}} {\overline  \oplus} W^t_{21,\{ 1,5 \}} {\overline  \oplus} W^t_{22,\{ 1,3 \}}, W^t_{21,\{ 4,5 \}} {\overline  \oplus} W^t_{32,\{ 1,5 \}} {\overline  \oplus} W^t_{22,\{ 1,5 \}},\nonumber\\
& W^t_{11,\{ 3,4 \}} {\overline  \oplus} W^t_{21,\{ 2,4 \}} {\overline  \oplus} W^t_{32,\{ 2,3 \}}, W^t_{11,\{ 3,5 \}} {\overline  \oplus} W^t_{21,\{ 2,5 \}} {\overline  \oplus} W^t_{22,\{ 2,3 \}},  W^t_{11,\{ 4,5 \}} {\overline  \oplus} W^t_{32,\{ 2,5 \}} {\overline  \oplus} W^t_{22,\{ 2,4 \}},\nonumber\\
& W^t_{21,\{ 4,5 \}} {\overline  \oplus} W^t_{32,\{ 3,5 \}} {\overline  \oplus} W^t_{22,\{ 3,4 \}}. 
\normalsize
\end{align}

\begin{itemize}
\item For $z=4$:
\end{itemize}
\begin{align}\label{DeliveredContentsAlg1z4Example}
\footnotesize
& W^t_{21,\{ 2,3,4 \}} {\overline  \oplus} W^t_{11,\{ 1,3,4 \}} {\overline  \oplus} W^t_{21,\{ 1,2,4 \}} {\overline  \oplus} W^t_{32,\{ 1,2,3 \}}, W^t_{21,\{ 2,3,5 \}} {\overline  \oplus} W^t_{11,\{ 1,3,5 \}} {\overline  \oplus} W^t_{21,\{ 1,2,5 \}} {\overline  \oplus} W^t_{22,\{ 1,2,3 \}}, \nonumber\\
& W^t_{21,\{ 2,4,5 \}} {\overline  \oplus} W^t_{11,\{ 1,4,5 \}} {\overline  \oplus} W^t_{32,\{ 1,2,5 \}} {\overline  \oplus} W^t_{22,\{ 1,2,4 \}}, W^t_{21,\{ 3,4,5 \}} {\overline  \oplus} W^t_{21,\{ 1,4,5 \}} {\overline  \oplus} W^t_{32,\{ 1,3,5 \}} {\overline  \oplus} W^t_{22,\{ 1,3,4 \}}, \nonumber\\
& W^t_{11,\{ 3,4,5 \}} {\overline  \oplus} W^t_{21,\{ 2,4,5 \}} {\overline  \oplus} W^t_{32,\{ 2,3,5 \}} {\overline  \oplus} W^t_{22,\{ 2,3,4 \}}.
\normalsize
\end{align}
\begin{itemize}
\item For $z=5$:
\end{itemize}
\begin{align}\label{DeliveredContentsAlg1z5Example}
\footnotesize
W^t_{21,\{ 2,3,4,5 \}} {\overline  \oplus} W^t_{11,\{ 1,3,4,5 \}} {\overline  \oplus} W^t_{21,\{ 1,2,4,5 \}} {\overline  \oplus} W^t_{32,\{ 1,2,3,5 \}} {\overline  \oplus} W^t_{22,\{ 1,2,3,4 \}}.
\normalsize
\end{align}

It can be seen that, having received the coded contents in \eqref{DeliveredContentsAlg1z1Example}$-$\eqref{DeliveredContentsAlg1z5Example}, each user $k$, $k \in [5]$, can recover the missing bits of its requested chunk, i.e., $\bigcup\nolimits_{\mathcal{S} \subset [5]: k \notin \mathcal{S}} W^t_{d_{k,t},\mathcal{S}}$.

Now we investigate the delivery scheme of Algorithm \ref{deliveryscheme}. The packets delivered for each part of the delivery phase are as follows:
\begin{itemize}
\item Part 1: 
\begin{align}\label{DeliveredContentsPart1Example}
\footnotesize
W^t_{21,\emptyset},W^t_{11,\emptyset},W^t_{32,\emptyset},W^t_{22,\emptyset}. 
\normalsize
\end{align}
\item Part 2.1: coded packets in \eqref{DeliveredContentsAlg1z2Example} are delivered.
\item Part 2.2: 
\begin{align}\label{DeliveredContentsPart22Example}
\footnotesize
& W^t_{21,\{ 1 \}} {\overline  \oplus} W^t_{21,\{ 2 \}}, W^t_{21,\{ 2 \}} {\overline  \oplus} W^t_{21,\{ 3 \}}, W^t_{21,\{ 3 \}} {\overline  \oplus} W^t_{21,\{ 4 \}},W^t_{21,\{ 4 \}} {\overline  \oplus} W^t_{21,\{ 5 \}},\nonumber\\ 
& W^t_{11,\{ 1 \}} {\overline  \oplus} W^t_{11,\{ 2 \}}, W^t_{11,\{ 2 \}} {\overline  \oplus} W^t_{11,\{ 3 \}}, W^t_{11,\{ 3 \}} {\overline  \oplus} W^t_{11,\{ 4 \}},W^t_{11,\{ 4 \}} {\overline  \oplus} W^t_{11,\{ 5 \}},\nonumber\\
& W^t_{32,\{ 1 \}} {\overline  \oplus} W^t_{32,\{ 2 \}}, W^t_{32,\{ 2 \}} {\overline  \oplus} W^t_{32,\{ 3 \}}, W^t_{32,\{ 3 \}} {\overline  \oplus} W^t_{32,\{ 4 \}},W^t_{32,\{ 4 \}} {\overline  \oplus} W^t_{32,\{ 5 \}},\nonumber\\
& W^t_{22,\{ 1 \}} {\overline  \oplus} W^t_{22,\{ 2 \}}, W^t_{22,\{ 2 \}} {\overline  \oplus} W^t_{22,\{ 3 \}}, W^t_{22,\{ 3 \}} {\overline  \oplus} W^t_{22,\{ 4 \}},W^t_{22,\{ 4 \}} {\overline  \oplus} W^t_{22,\{ 5 \}}.
\normalsize
\end{align}
\item Part 3: 
\begin{itemize}
\item For $\mathcal{P} \subset [5]$, and $\left| \mathcal{P} \right| = 3$: coded packets in \eqref{DeliveredContentsAlg1z3Example} are delivered.
\item For $\mathcal{P} \subset [5]$, and $\left| \mathcal{P} \right| = 4$: coded packets in \eqref{DeliveredContentsAlg1z4Example} are delivered.
\item For $\mathcal{P} \subset [5]$, and $\left| \mathcal{P} \right| = 5$: coded packets in \eqref{DeliveredContentsAlg1z5Example} are delivered.
\end{itemize}
\end{itemize}
Note that the total sizes of the coded packets in \eqref{DeliveredContentsAlg1z2Example} and \eqref{DeliveredContentsPart22Example} depend on the cache content distribution $\mathbf{Q}$, and among Part 2.1 and Part 2.2 the one that delivers less number of bits in total is executed. Part 1 of Algorithm \ref{deliveryscheme} is obviously more efficient than the bits delivered for $z=1$ in Algorithm \ref{deliveryscheme1} as the former removes the repeated bits in the latter. It can be seen that the user demanding chunk $W_{ij}\in \mathcal{D}_t$, $i \in [3]$ and $j \in [2]$, can obtain $\bigcup\nolimits_{\mathcal{S} \subset [5]: \left| \mathcal{S} \right| = 1} W^t_{ij,\mathcal{S}}$ after receiving $\bigcup\nolimits_{k =1}^{4}W^t_{ij, \{k\}}\overline{\oplus}W^t_{i j,\{k+1\}}$, delivered in Part 2.2 of Algorithm \ref{deliveryscheme}, thanks to its cache content. For example, user 3, which requests chunk $W_{21}$ and has access to $W^t_{21,\{ 3 \}}$ locally, can obtain $\bigcup\nolimits_{\mathcal{S} \subset [5]: \left| \mathcal{S} \right| = 1} W^t_{21,\mathcal{S}}$ after receiving
\begin{align}\label{DeliveredContentsUser3Part22Example}
W^t_{21,\{ 1 \}} {\overline  \oplus} W^t_{21,\{ 2 \}}, W^t_{21,\{ 2 \}} {\overline  \oplus} W^t_{21,\{ 3 \}}, W^t_{21,\{ 3 \}} {\overline  \oplus} W^t_{21,\{ 4 \}},W^t_{21,\{ 4 \}} {\overline  \oplus} W^t_{21,\{ 5 \}}.
\end{align}
The coded packets delivered in Part 3 of Algorithm \ref{deliveryscheme} are the same as those delivered by Algorithm \ref{deliveryscheme1} for $z \ge 3$; and they enable user $k$ to recover $\bigcup\nolimits_{\mathcal{S} \subset [5]: \left| \mathcal S \right| \ge 2, k \notin \mathcal{S}} W^t_{d_{k,t},\mathcal{S}}$. Thus, together with their cache contents, all the users can recover their demands. 

\subsection{Average Delivery Rate}\label{sec:rate}
Here we present a closed-form expression for the achievable average delivery rate of the proposed coded caching scheme. For ease of presentation, we first introduce some notations:

\begin{itemize}
    \item Let $p^j$ denote the probability that a user watches the $j$th chunk. We have
    \begin{equation}
    p^j=\sum\nolimits_{i=1}^N p_ip_{ij}, \quad \forall j \in [B]. 
    \end{equation}
    \item For any time slot, let $\mathcal{S}_j$, for $j \in [B]$, denote the set of users requesting their $j$th chunks, while $\mathcal{S}_{\mathrm{all}} = \bigcup\nolimits_{j=1}^B \mathcal{S}_{j}$ is the set of all active users. Letting $K_j \triangleq \left| \mathcal{S}_j \right|$, for $j \in [B]$, we have\footnote{Note that we remove the dependency of $K^{(t)}_j$ on $t$ and replace it by $K_j$ to make the notation valid for any time slot.}
    \begin{equation}
    \mathrm{Pr}\{K_j=k\}=\sum\nolimits_{a=k}^{A_{max}} P_A(a)\binom{a}{k}(p^j)^{k}(1-p^j)^{a-k}, \quad \forall k \in \mathcal{A}, j \in [B], 
    \end{equation}
    where we assume $0^0=1$.
% (note that we denoted by $K^{(t)}_j$ the number of users requesting $j$th chunk at time slot $t$ in the previous section. Here to make the notation valid for any time slot, we remove the dependency on time $t$)    
    \item Let $\tilde{p}_{ij}$ denote the probability of a user requesting $W_{ij}$ given that she is demanding the $j$th chunk of a file, i.e., $\tilde{p}_{ij}=p_ip_{ij}/p^j$, $\forall i \in [N], j \in [B]$. Note that we have $\sum\nolimits_{i=1}^N\tilde{p}_{ij}=1$. We refer to $\tilde{p}_{ij}$ as the \textit{normalized popularity} of chunk $W_{ij}$, for $i \in [N]$ and $j \in [B]$. 
    \item For a given set of $l \ge 1$ users, we define $g_{ij, (l, l')}$ as the number of bits of chunk $W_{ij}$, normalized by $F/B$, that have been cached by a subset of $l'$ users among the $l$ users, and not by any of the remaining $l-l'$ users, for $l' \in [l]$. Due to the law of large numbers, $g_{ij, (l, l')}$ is identical for any $l$ users and any $l'$ users among them, and we have  
    \begin{equation}
    g_{ij, (l, l')} = (q_{ij})^{l'}(1-q_{ij})^{l-l'}, \quad  \forall i \in [N], j \in [B],
    \end{equation}
    with probability 1 as $F \to \infty$. Recall that $q_{ij}$ is the caching probability for chunk $W_{ij}$ as defined in Section~\ref{placement}, which is identical across users. Let $\mathcal{F}_{ij}\triangleq \{W_{fh}, f \in [N], h \in [B]: q_{fh}=q_{ij}\}$, for $i \in [N]$ and $j \in [B]$; that is the set of chunks for which an independent random subset of $q_{ij}F/B$ bits of them are cached by each user. We note that $W_{ij} \in \mathcal{F}_{ij}$. We also note that $\forall W_{fh} \in \mathcal{F}_{ij}$, $ g_{fh, (l, l')}=g_{ij, (l, l')}$, $\forall l, l'$. 
    \item For a given time slot, let $\mathcal{S}'_{j}$ be an $l_j$-element subset of $\mathcal{S}_{j}$, $j\in [B]$, and $\mathcal{S}_{\mathrm{sub}}\triangleq \bigcup\nolimits_{j=1}^B \mathcal{S}'_{j}$, $\mathcal{S}_{\mathrm{all}}\triangleq \bigcup\nolimits_{j=1}^B \mathcal{S}_{j}$. We denote by $\mathcal{D}_{\mathcal{S}_{\mathrm{sub}}}$ the demand combination of the users in $\mathcal{S}_{\mathrm{sub}}$, $\mathcal{D}_{\mathcal{S}_{\mathrm{sub}}}\in \mathfrak{D}_{l_{[B]}}$,  $\mathfrak{D}_{l_{[B]}}\triangleq\{W_{11}, ..., W_{N1}\}^{l_1}\times \{W_{12}, ..., W_{N2}\}^{l_2}\times \cdots \times \{W_{1B}, ..., W_{NB}\}^{l_B}$. Let
    \begin{align}\label{defi:rho}
   \rho_{ij, (\mathcal{S}_{\mathrm{all}}, \mathcal{S}_{\mathrm{sub}})}\triangleq \mathrm{Pr}\bigg\{\max\limits_{W_{fh} \in \mathcal{D}_{\mathcal{S}_{\mathrm{sub}}}} g_{fh, (\sum\nolimits_{s=1}^{B}K_s, \sum\nolimits_{s=1}^{B}l_s-1)}=g_{ij, (\sum\nolimits_{s=1}^{B}K_s, \sum\nolimits_{s=1}^{B}l_s-1)}\bigg\},
\end{align}
 $\forall i\in [N],j\in [B]$, that is, $\rho_{ij, (\mathcal{S}_{\mathrm{all}}, \mathcal{S}_{\mathrm{sub}})}$ is the probability that the maximum number of bits of a requested chunk by $\mathcal{S}_{\mathrm{sub}}$ cached exclusively by $\sum\nolimits_{s=1}^{B}l_s-1$ users in $\mathcal{S}_{\mathrm{sub}}$ (and not cached by the rest of the users in $\mathcal{S}_{\mathrm{all}}$), which is identical for any $\sum\nolimits_{s=1}^{B}l_s-1$ users in $\mathcal{S}_{\mathrm{sub}}$, is given by $g_{ij, \left( \sum\nolimits_{s=1}^{B}K_s, \sum\nolimits_{s=1}^{B}l_s-1 \right)}$. Since the file popularity and audience retention rates are identical among the users, the distribution of $\mathcal{D}_{ \mathcal{S}_{\mathrm{sub}}}$ only depends on $l_{[B]}$. Thus, for simplicity, we use $\mathcal{D}_{l_{[B]}}$ and $\rho_{ij, (K_{[B]},l_{[B]})}$ instead of $\mathcal{D}_{\mathcal{S}_{\mathrm{sub}}}$ and $\rho_{ij, (\mathcal{S}_{\mathrm{all}},\mathcal{S}_{\mathrm{sub}})}$, respectively.

 \end{itemize}

%Next we present the average delivery rate achieved by the MAN and PCC schemes in Theorem~\ref{theorem:rate1} and Theorem~\ref{theorem:rate}, respectively. 

\begin{Theorem}\label{theorem:rate0}
For the caching system described in Section~\ref{system}, and a given cache content distribution $\mathbf{Q}$, the following average delivery rate is achieved by the placement scheme presented in Section~\ref{placement} followed by the $\rm{RAN}$ delivery scheme presented in Algorithm~\ref{randomdelivery}:
\begin{equation}
\begin{aligned}\label{m1}
R_{\rm{RAN}}(P_A, \mathbf{p}, \mathbf{P}, \mathbf{Q})=\sum\nolimits_{j=1}^B \sum\nolimits_{i=1}^N\sum\nolimits_{k=0}^{A_{max}}\mathrm{Pr}\{K_j=k\} \Big(1-\left(1-\tilde{p}_{ij}\right)^K\Big)(1-q_{ij}).
\end{aligned}
\end{equation}
\begin{proof}
The detailed proof can be found in Appendix A.
\end{proof}
\end{Theorem}
\begin{Remark}
Given $\mathbf{Q}$, consider the uncoded caching and delivery scheme, shortly referred to as \textit{Uncoded}, where each user caches the same $q_{ij}F/B$ bits from chunk $W_{ij}$, $i\in [N]$, $j\in [B]$, in the placement phase. At each time slot of the delivery phase, the server sends the missing  $(1-q_{ij})F/B$ bits of chunk $W_{ij}$ if it is requested. We note that for any demand combination, the Uncoded scheme sends the same number of bits as the $\rm{RAN}$ delivery scheme, for the placement scheme described in Section~\ref{placement}, which results in the same average delivery rate given in \eqref{m1}.   
\end{Remark}

\begin{Theorem}\label{theorem:rate1}
For the caching system described in Section~\ref{system}, and a given cache content distribution $\mathbf{Q}$, the following average delivery rate is achieved by the placement scheme presented in Section~\ref{placement} followed by the $\rm{MAN}$ delivery scheme presented in Algorithm~\ref{deliveryscheme1}:
\begin{align}\label{varphi1}
&R_{\rm{MAN}}(P_A, \mathbf{p}, \mathbf{P}, \mathbf{Q})=\sum\nolimits_{k_{[B]}\in \mathcal{A}^B}\left(\prod\nolimits_{s=1}^B \mathrm{Pr}\{K_s=k_s\}\right)\nonumber\\ 
&\quad \left(\sum\nolimits_{l_{[B]}\in [0:k_1]\times\cdots\times [0:k_B]}\prod\nolimits_{j=1}^B\binom{k_j}{l_j}\sum\nolimits_{j=1}^B \sum\nolimits_{i=1}^N \rho'_{ij, (k_{[B]},~l_{[B]})}g_{ij, (\sum\nolimits_{s=1}^B k_s, \sum\nolimits_{s=1}^B l_s-1)}\right),
\end{align}
where
\begin{equation}\label{defi:rhoprime}
\rho'_{ij, (k_{[B]},~l_{[B]})}\triangleq\frac{\rho_{ij, (k_{[B]},~l_{[B]})}}{\sum\nolimits_{f=1}^N\sum\nolimits_{h=1}^B \mathbbm{1}\left\{g_{fh, (\sum\nolimits_{s=1}^{B}k_s, \sum\nolimits_{s=1}^{B}l_s)}=g_{ij, (\sum\nolimits_{s=1}^{B}k_s, \sum\nolimits_{s=1}^{B}l_s)}\right\}}.   
\end{equation}
\begin{proof}
The detailed proof can be found in Appendix B.
\end{proof}
\end{Theorem}
\begin{Remark}
We remark that when $B=1$ and $\mathcal{A}=[K]$, the considered caching problem reduces to the one with non-uniform file popularities~\cite{JiArXivNonuniform}. Rate $\min\{R_{\rm{RAN}}(P_A, \mathbf{p}, \mathbf{P}, \mathbf{Q}), R_{\rm{MAN}}(P_A, \mathbf{p}, \mathbf{P}, \mathbf{Q})\}$ can be achieved by performing the scheme resulting in a smaller delivery rate among the $\rm{RAN}$ and $\rm{MAN}$ schemes for given $\mathbf{Q}$. For a given $\mathbf{Q}$,  where $q_{ij}=q_{fh}$ for some $ij\neq fh, i, f \in [N], j, h \in [B]$, (such that $g_{fh, \left(\sum\nolimits_{s=1}^{B}k_s, \sum\nolimits_{s=1}^{B}l_s \right)}=g_{ij, \left(\sum\nolimits_{s=1}^{B}k_s, \sum\nolimits_{s=1}^{B}l_s\right)}$ for any $k_{[B]}$ and $l_{[B]}$), it provides a tighter upper bound compared to the one characterized in \cite[Theorem 1]{JiArXivNonuniform} due to the denominator in \eqref{defi:rhoprime}. However, the optimization of cache allocation over the average delivery rate will ensure that the cache capacities allocated to different files are distinct. Hence, with optimal cache allocation the upper bound in \cite[Theorem 1]{JiArXivNonuniform} can be arbitrarily close to $\min\{R_{\rm{RAN}}(P_A, \mathbf{p}, \mathbf{P}, \mathbf{Q}), R_{\rm{MAN}}(P_A, \mathbf{p}, \mathbf{P}, \mathbf{Q})\}$. 
\end{Remark}

\begin{Theorem}\label{theorem:rate}
For the caching system described in Section~\ref{system}, and a given cache content distribution $\mathbf{Q}$, the following average delivery rate is achieved by the placement scheme presented in Section~\ref{placement} followed by the $\rm{PCC}$ delivery scheme outlined in Algorithm~\ref{deliveryscheme}:
\begin{subequations}
\begin{align}
&R_{\rm{PCC}}(P_A, \mathbf{p}, \mathbf{P}, \mathbf{Q})\triangleq R_{\rm{MAN}}(P_A, \mathbf{p}, \mathbf{P}, \mathbf{Q})-\Delta \varphi_1(P_A, \mathbf{p}, \mathbf{P}, \mathbf{Q})-\Delta \varphi_2(P_A, \mathbf{p}, \mathbf{P}, \mathbf{Q}),\label{varphi}\\ 
&\Delta \varphi_1(P_A, \mathbf{p}, \mathbf{P}, \mathbf{Q})\triangleq \sum\nolimits_{k_{[B]}\in \mathcal{A}^B}\left(\prod\nolimits_{s=1}^B \mathrm{Pr}\{K_s=k_s\}\right)\nonumber\\ 
&\quad \left(\sum\nolimits_{j=1}^Bk_j\sum\nolimits_{i=1}^N\tilde{p}_{ij}g_{ij, (\sum\nolimits_{s=1}^B k_s, 0)}- \sum\nolimits_{j=1}^B \sum\nolimits_{i=1}^N \left(1-\left(1-\tilde{p}_{ij}\right)^{k_j}\right)g_{ij, (\sum\nolimits_{s=1}^B k_s, 0)}\right)\label{deltavarphi1},\\
&\Delta \varphi_2(P_A, \mathbf{p}, \mathbf{P}, \mathbf{Q}) \triangleq \sum\nolimits_{k_{[B]}\in \mathcal{A}^B}\left(\prod\nolimits_{s=1}^B \mathrm{Pr}\{K_s=k_s\}\right)\nonumber\\ &\qquad\qquad\max\left\{ \sum\nolimits_{j=1}^B\binom{k_j}{2}\sum\nolimits_{i=1}^N\rho'_{ij, (k_{[B]},~(0, ..., l_j=2, ..., 0))} g_{ij, (\sum\nolimits_{s=1}^B k_s, 1)} \right.\nonumber\\
&\qquad\qquad +\sum\nolimits_{j_1=1}^B k_{j_1}\sum\nolimits_{j_2=j_1+1}^B k_{j_2}\sum\nolimits_{i=1}^N\sum\nolimits_{j=1}^B\rho'_{ij, (k_{[B]},~(0, ..., l_{j_1}=1, ..., l_{j_2}=1, ..., 0))}g_{ij, (\sum\nolimits_{s=1}^B k_s, 1)}\nonumber\\
&\qquad\qquad\left.-\sum\nolimits_{j=1}^B \sum\nolimits_{i=1}^N(\sum\nolimits_{s=1}^B k_s-1) \left(1-\left(1-\tilde{p}_{ij}\right)^{k_{j}}\right)g_{ij, (\sum\nolimits_{s=1}^B k_s, 1)}, 0\right\},\label{deltavarphi2}
\end{align}
\end{subequations} 
where $(0, ..., l_j=2, ..., 0)$ is a $B$-element vector, such that the $j$th element is $2$ and all the other elements are zero, for some $j \in [B]$, while $(0, ..., l_{j_1}=1, ..., l_{j_2}=1, ..., 0)$ is a $B$-element vector, such that the $j_1$th and $j_2$th elements are $1$ while the rest are zero, for some $j_1, j_2 \in [B]$, $j_1 \ne j_2$. 
\begin{proof}
Observe that the missing bits sent in PART 1 and PART 2 of Algorithm \ref{deliveryscheme} are sent by the delivery scheme in Algorithm \ref{deliveryscheme1} for $z=1$ and $z=2$, respectively; the messages sent in PART 3 are the same messages sent by the delivery scheme in Algorithm \ref{deliveryscheme1} for $z>2$. We point out here that $\Delta \varphi_1(P_A, \mathbf{p}, \mathbf{P}, \mathbf{Q})$ is the difference between the average number of bits sent by PART 1 of Algorithm~\ref{deliveryscheme} and those sent by the delivery scheme in Algorithm~\ref{deliveryscheme1} for $z=1$, while $\Delta \varphi_2(P_A, \mathbf{p}, \mathbf{P}, \mathbf{Q})$ is the difference between the average number of bits sent by PART 2 of Algorithm~\ref{deliveryscheme} and those sent by the delivery scheme in Algorithm~\ref{deliveryscheme1} for $z=2$. Hence, we have the delivery rate achieved by the delivery scheme in Algorithm~\ref{deliveryscheme} as in \eqref{varphi}. The detailed proofs of \eqref{deltavarphi1} and \eqref{deltavarphi2} can be found in Appendix C. 
\end{proof}
\end{Theorem}
The value of $\rho_{ij, (k_{[B]},l_{[B]})}$ can be calculated as follows. We define, $\forall D_{l_{[B]}}\in \mathfrak{D}_{l_{[B]}}$,   
\begin{equation}
Y_{k_{[B]},l_{[B]}}(D_{l_{[B]}})\triangleq \max\limits_{W_{fh} \in D_{l_{[B]}}}g_{fh, (\sum\nolimits_{s=1}^{B}k_s, \sum\nolimits_{s=1}^{B}l_s-1)}.
\end{equation}
Let
\begin{equation}
\mathfrak{D}'_{l_{[B]}, ij}\triangleq \left\{D_{l_{[B]}}\in \mathfrak{D}_{l_{[B]}}:  Y_{k_{[B]},l_{[B]}}(D_{l_{[B]}}) \leq g_{ij, (\sum\nolimits_{s=1}^{B}k_s, \sum\nolimits_{s=1}^{B}l_s-1)}\right\}, \nonumber  \end{equation}
i.e., $\mathfrak{D}'_{l_{[B]}, ij}$ is the set of all elements $D_{l_{[B]}}$ in $\mathfrak{D}_{l_{[B]}}$ such that $Y_{k_{[B]},l_{[B]}}(D_{l_{[B]}}) \leq g_{ij, (\sum\nolimits_{s=1}^{B}k_s, \sum\nolimits_{s=1}^{B}l_s-1)}$. Similarly, let 
\begin{equation}
\mathfrak{D}''_{l_{[B]}, ij}\triangleq \left\{D_{l_{[B]}}\in \mathfrak{D}_{l_{[B]}}:  Y_{k_{[B]},l_{[B]}}(D_{l_{[B]}}) < g_{ij, (\sum\nolimits_{s=1}^{B}k_s, \sum\nolimits_{s=1}^{B}l_s-1)}\right\}, \nonumber  \end{equation}
and 
\begin{equation}
\mathfrak{D}'''_{l_{[B]}, ij}\triangleq \left\{D_{l_{[B]}}\in \mathfrak{D}_{l_{[B]}}:  Y_{k_{[B]},l_{[B]}}(D_{l_{[B]}})=g_{ij, (\sum\nolimits_{s=1}^{B}k_s, \sum\nolimits_{s=1}^{B}l_s-1)}\right\}. \nonumber  \end{equation}
We have $\mathfrak{D}'''_{l_{[B]}, ij}=\mathfrak{D}'_{l_{[B]}, ij}\setminus\mathfrak{D}''_{l_{[B]}, ij}$. It follows that
\begin{equation}\label{y1}
\begin{aligned}
\sum\limits_{D_{l_{[B]}}\in\mathfrak{D}'_{l_{[B]}, ij}}\mathrm{Pr}&\left\{\mathcal{D}_{l_{[B]}}=D_{l_{[B]}}\right\}=\prod\limits_{h=1}^B\left(\sum\nolimits_{W_f\in \mathcal{F}:g_{fh, (\sum\nolimits_{s=1}^{B}k_s, \sum\nolimits_{s=1}^{B}l_s-1)}\leq g_{ij, (\sum\nolimits_{s=1}^{B}k_s, \sum\nolimits_{s=1}^{B}l_s-1)}}\tilde{p}_{fh}\right)^{l_h}, 
\end{aligned}
\end{equation}
that is, the probability that a demand combination $\mathcal{D}_{l_{[B]}}$ falls in the set $\mathfrak{D}'_{l_{[B]}, ij}$, i.e., $Y_{k_{[B]},l_{[B]}}(\mathcal{D}_{l_{[B]}})$ $\leq g_{ij, (\sum\nolimits_{s=1}^{B}k_s, \sum\nolimits_{s=1}^{B}l_s-1)}$, is the probability that each requested chunk, $W_{fh}\in \mathcal{D}_{l_{[B]}}$, is associated with $g_{fh, (\sum\nolimits_{s=1}^{B}k_s, \sum\nolimits_{s=1}^{B}l_s-1)}$ no greater than $g_{ij, (\sum\nolimits_{s=1}^{B}k_s, \sum\nolimits_{s=1}^{B}l_s-1)}$, given that there are $l_h$ requests of the $h$-th chunks, $\forall h\in [B]$. Similarly,
\begin{equation}\label{y2}
\begin{aligned}
\sum\limits_{D_{l_{[B]}}\in\mathfrak{D}''_{l_{[B]}, ij}}\mathrm{Pr}&\left\{\mathcal{D}_{l_{[B]}}=D_{l_{[B]}}\right\}=\prod\limits_{h=1}^B\left(\sum\nolimits_{W_f\in \mathcal{F}:g_{fh, (\sum\nolimits_{s=1}^{B}k_s, \sum\nolimits_{s=1}^{B}l_s-1)}< g_{ij, (\sum\nolimits_{s=1}^{B}k_s, \sum\nolimits_{s=1}^{B}l_s-1)}}\tilde{p}_{fh}\right)^{l_h}, 
\end{aligned}
\end{equation}
i.e., the probability that the value of $g_{fh, (\sum\nolimits_{s=1}^{B}k_s, \sum\nolimits_{s=1}^{B}l_s-1)}$ for each requested chunk, $W_{fh}\in \mathcal{D}_{l_{[B]}}$, is less than $g_{ij, (\sum\nolimits_{s=1}^{B}k_s, \sum\nolimits_{s=1}^{B}l_s-1)}$, given that there are $l_h$ requests of the $h$-th chunks, $\forall h\in [B]$ (note that it is ``no larger than'' in the case of \eqref{y1}). According to the definition of $\rho_{ij, (k_{[B]},l_{[B]})}$ given in \eqref{defi:rho} and the fact that $\mathfrak{D}'''_{l_{[B]}, ij}\equiv \mathfrak{D}_{l_{[B]}, ij}$, we derive 
\begin{align}\label{rho}
&\rho_{ij, (k_{[B]},~l_{[B]})} =\sum\nolimits_{D_{\mathcal{S}_{\mathrm{sub}}}\in\mathfrak{D}'''_{\mathcal{S}_{\mathrm{sub}}, ij} }\mathrm{Pr}\bigg\{\mathcal{D}_{\mathcal{S}_{\mathrm{sub}}}=D_{\mathcal{S}_{\mathrm{sub}}}\bigg\}\nonumber\\
& \qquad =\sum\nolimits_{D_{\mathcal{S}_{\mathrm{sub}}}\in\mathfrak{D}'_{\mathcal{S}_{\mathrm{sub}}, ij} }\mathrm{Pr}\bigg\{\mathcal{D}_{\mathcal{S}_{\mathrm{sub}}}=D_{\mathcal{S}_{\mathrm{sub}}}\bigg\}-\sum\nolimits_{D_{\mathcal{S}_{\mathrm{sub}}}\in\mathfrak{D}''_{\mathcal{S}_{\mathrm{sub}}, ij} }\mathrm{Pr}\bigg\{\mathcal{D}_{\mathcal{S}_{\mathrm{sub}}}=D_{\mathcal{S}_{\mathrm{sub}}}\bigg\}.
\end{align}
Thus, $\rho_{ij, (k_{[B]},~l_{[B]})}$ can be easily calculated by sorting $\left\{g_{fh, (\sum\nolimits_{s=1}^{B}k_s, \sum\nolimits_{s=1}^{B}l_s -1)}, f\in [N], h \in [B]\right\}$.
\begin{Remark}
It is trivial to see that $\Delta \varphi_1(P_A, \mathbf{p}, \mathbf{P}, \mathbf{Q})\geq 0$, and the equality holds only when $\mathcal{A}=\{1\}$, and $\Delta \varphi_2(P_A, \mathbf{p}, \mathbf{P}, \mathbf{Q})\geq 0$ according to \eqref{deltavarphi2}. Hence, we can conclude that the PCC scheme achieves a lower average delivery rate than the $\rm{MAN}$ scheme when $\mathcal{A}\neq\{1\}$, which will further be validated by the numerical results. 
\end{Remark}

\subsection{Cache Allocation}\label{cache}
We formulate the optimization of the cache content distribution $\mathbf{Q}$ as follows: 
\begin{subequations}\label{optimization}
\begin{align}
&\min R(\mathbf{Q}) \label{object1}\\
&\mathrm{s. t.} \sum\nolimits_{i,j}q_{ij}=MB,\label{constrain}
\end{align}
\end{subequations}
where the objective is to minimize the average delivery rate over all possible demand combinations while the cache capacity constraint at each user is satisfied with equality. We consider $R_{\rm{MAN}}(\mathbf{Q})$ and  $R_{\rm{PCC}}(\mathbf{Q})$ as the objective function, for the MAN and PCC schemes, respectively. The optimization problem in \eqref{optimization} will be solved numerically, and the corresponding solution will be referred to as the \textit{optimal cache allocation} (OCA).

However, in practice, there will be a large number of files in the library, and each video file can be partitioned into many chunks. In that case, optimizing $\mathbf{Q}$ over all the chunks in the library requires high computational complexity. As an alternative, we present a low-complexity cache allocation scheme, referred to as \textit{popularity based cache allocation} (PCA), in which only the most popular chunks are cached by the users; that is, we have
\begin{align}\label{LRU} q_{ij}=\begin{cases}
q,&\mbox{if $p_ip_{ij}\geq \overline{n}$},\\
0,&\mbox{otherwise},
\end{cases}
\end{align}
where $q \in (0, 1]$ and $\overline{n}$ are the two  parameters to be chosen to satisfy $\sum\nolimits_{i=1}^N\sum\nolimits_{j=1}^Bq_{ij}=MB$. We denote the cache content distribution given by \eqref{LRU} as a function of $q$, i.e., $\mathbf{Q}(q)$. The optimization of $q$ can be expressed as $q^* \buildrel \Delta \over = \argmin R(\mathbf{Q}(q))\label{object2}$, which can be computed through one-dimensional search. The comparison of the presented caching schemes through numerical simulations is relegated to Section \ref{section:numerical}.

\section{Lower Bound}\label{SecLowerBound}
Here, we present a lower bound on the average delivery rate-cache capacity tradeoff $R(M)$, derived by assuming that some requested chunks are served by a genie at no transmission cost. In this way the problem is relaxed to a caching problem with uniform file popularity, whose delivery rate can be bounded using cut-set arguments. We note, however, that the derived lower bound is loose due to this relaxation. 
\begin{Theorem}\label{lowerbound}
For the caching problem described in Section~\ref{system}, $R(M)$ is lower bounded by
\begin{align}
R(M)\geq R^* (M) \triangleq & \sum\limits_{k_{[B]}\in \mathcal{A}^B}\left(\prod\limits_{j=1}^B \mathrm{Pr}\{K_j=k_j\}\right)\operatorname*{max}\limits_{n_{[B]},v_{[B]}, \widetilde{z}_{[B]}}\left\{\left(\prod\limits_{j=1}^Bf'_j(k_j, n_j, v_j)\right)\cdot\right.\nonumber\\
&\quad\left.\left(\prod\limits_{j=1}^Bf''_j(n_j, v_j, \widetilde{z}_j)\right)\operatorname*{max}\limits_{\substack{z_j\in [\lceil \min\{\widetilde{z}_j, v_j\}\rceil]\\j \in [B]}}\left\{\sum\limits_{j=1}^B z_j \left(1-\frac{MB}{\min\limits_{j\in [B]}\lfloor\frac{n_j}{z_j}\rfloor}\right)\right\}\right\},
\end{align}
where $n_j \in [N]$, $v_j\in (0, k_jn_jr_{n_jj}]$, $\widetilde{z}_{j}\in (0, f(n_j, v_j)]$, $j\in [B]$, and
\begin{equation}
f(n_j, v_j)\triangleq n_j\left(1-\left(1-\frac{1}{n_j}\right)^{v_j}\right),
\end{equation}
\begin{equation}
f'_j(k_j, n_j, v_j)\triangleq 1-\exp\left(-\frac{(k_jn_jr_{n_jj}-v_j)^2}{2k_jn_jr_{n_jj}}\right),
\end{equation}
\begin{equation}
f''_j(n_j, v_j, z_j)\triangleq 1-\exp\left(-\frac{(f(n_j, v_j)-\widetilde{z}_j)^2}{2f(n_j, v_j)}\right),
\end{equation}
and $r_{1j}, ..., r_{Nj}$ is an ordered permutation of $\{\tilde{p}_{1j}, ..., \tilde{p}_{Nj}\}$, such that $r_{1j}\geq \cdots \geq r_{Nj}$, $\forall j \in [B]$. 
\begin{proof}
Here, $n_j$ is a parameter that controls which chunk is served by the genie. In particular, considering user $k$ demands $j$-th chunk of his requests, if the requested chunk has a normalized popularity  lower than $r_{n_jj}$, i.e.,  $\tilde{p}_{d_{k, t}}<r_{n_jj}$, it is served by a genie at no transmission cost; otherwise, i.e., if $\tilde{p}_{d_{k, t}}\geq r_{n_jj}$, it is served by a genie with probability $1-r_{n_jj}/\tilde{p}_{d_{k, t}}$; that is, the server has to transmit the required $j$-th chunk to this user through the shared link with probability $r_{n_jj}/\tilde{p}_{d_{k, t}}$. Given $k_j$ users requesting $j$-th chunks and parameter $n_j$, $f'_j(k_j, n_j, v_j)$ represents the probability of $v_j$ users among $k_j$ requiring service from the server, and the rest $k_j-v_j$ users served by the genie. $f''_j(n_j, v_j, z_j)$ denotes the probability that $z_j$ distinct $j$-th chunks are requested by these $v_j$ users who are requiring service from the server given parameter $n_j$. The detailed proof can be found in Appendix D.
\end{proof}
\end{Theorem}

\section{Numerical Results}\label{section:numerical}

\begin{figure}[!t]
\centering
%width=1.05\linewidth
\includegraphics[scale=0.6,trim={34pt 37pt 62pt 32pt},clip]{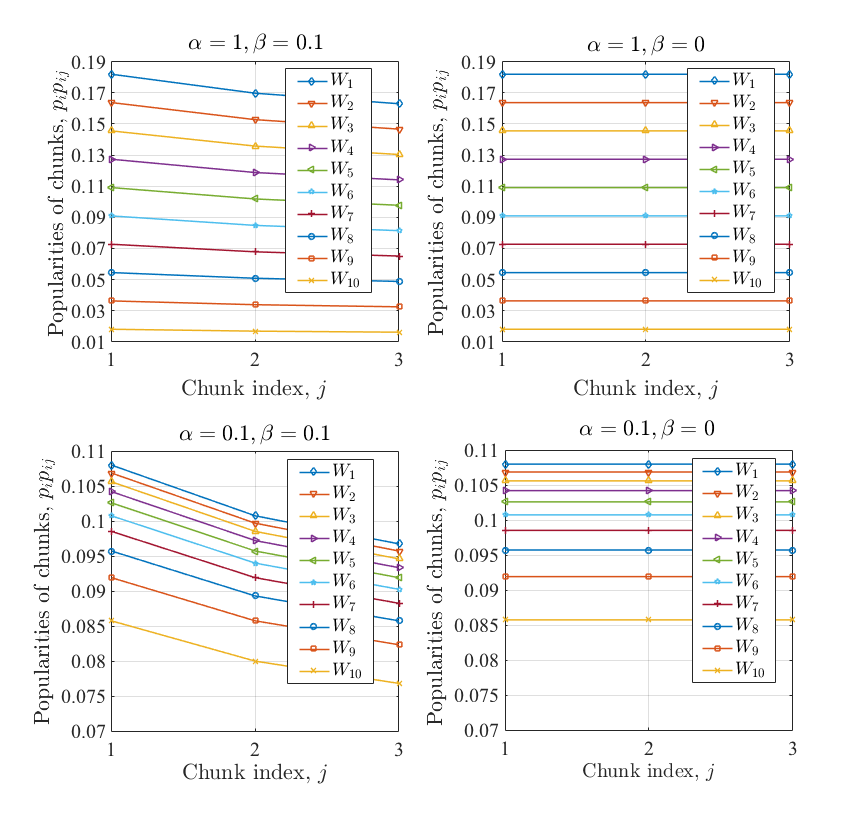}
%trim={left bottom right top}
\caption{The popularity of video chunks $W_{ij}$, i.e., $p_{i}p_{ij}$ given different values of $\alpha$ and $\beta$.}\label{reten}
\end{figure}

%\begin{figure}[t!]\label{fig:1}
%\centering
%\includegraphics[width=1\linewidth]{retention1.png}
%\caption{The popularity of video chunks $W_{ij}$, i.e., $p_{i}p_{ij}$ given $\beta_i=0.5-0.1i$, $i=1, ..., 5$, $j=1, ..., 5$.}
%\end{figure}

In this section, we numerically evaluate the average delivery rate achieved by the two coded delivery schemes, i.e., MAN and PCC, with both cache allocation strategies, OCA and PCA, and compare with RAN, the uncoded caching as well as the lower bound. In uncoded caching, each user fully caches as many of the most popular chunks as possible to fill its cache capacity. We consider $N=10$ video files in the library, each consists of $B=3$ chunks of equal size. We assume that the popularity of files follows a Zipf power law with parameter $\alpha$\cite{breslau1999web}, in which case we have $p_i=(6-i)^{\alpha}/\left({\sum\nolimits_{f=1}^N f^{\alpha}}\right)$, for $i\in [5]$, and the audience retention rates of the video files follow a Zipf-like distribution as well \cite{yu2006dynamic}, i.e., $p_{ij}=j^{-\beta_{i}}$, for $i\in [5],j\in [3]$, with parameter $\beta_{i}\geq0$. The larger $\beta_{i}$ implies a shorter average watching time for file $W_i$. We set $\beta_{i}$ to be identical for all the files, i.e., $\beta_{i}=\beta$, $\forall i\in[N]$, and the corresponding popularity of chunks, i.e., $p_ip_{ij}$, $i\in [5]$, $j\in [3]$, are presented in Fig. \ref{reten} for different values of $\alpha$ and $\beta$. It shows that a higher $\alpha$ results in a larger difference in file popularities. Moreover, with a larger $\beta$, the retention rates of the video files decrease more quickly with the chunk index.

\begin{figure}[!t]
\centering
\includegraphics[scale=0.34]{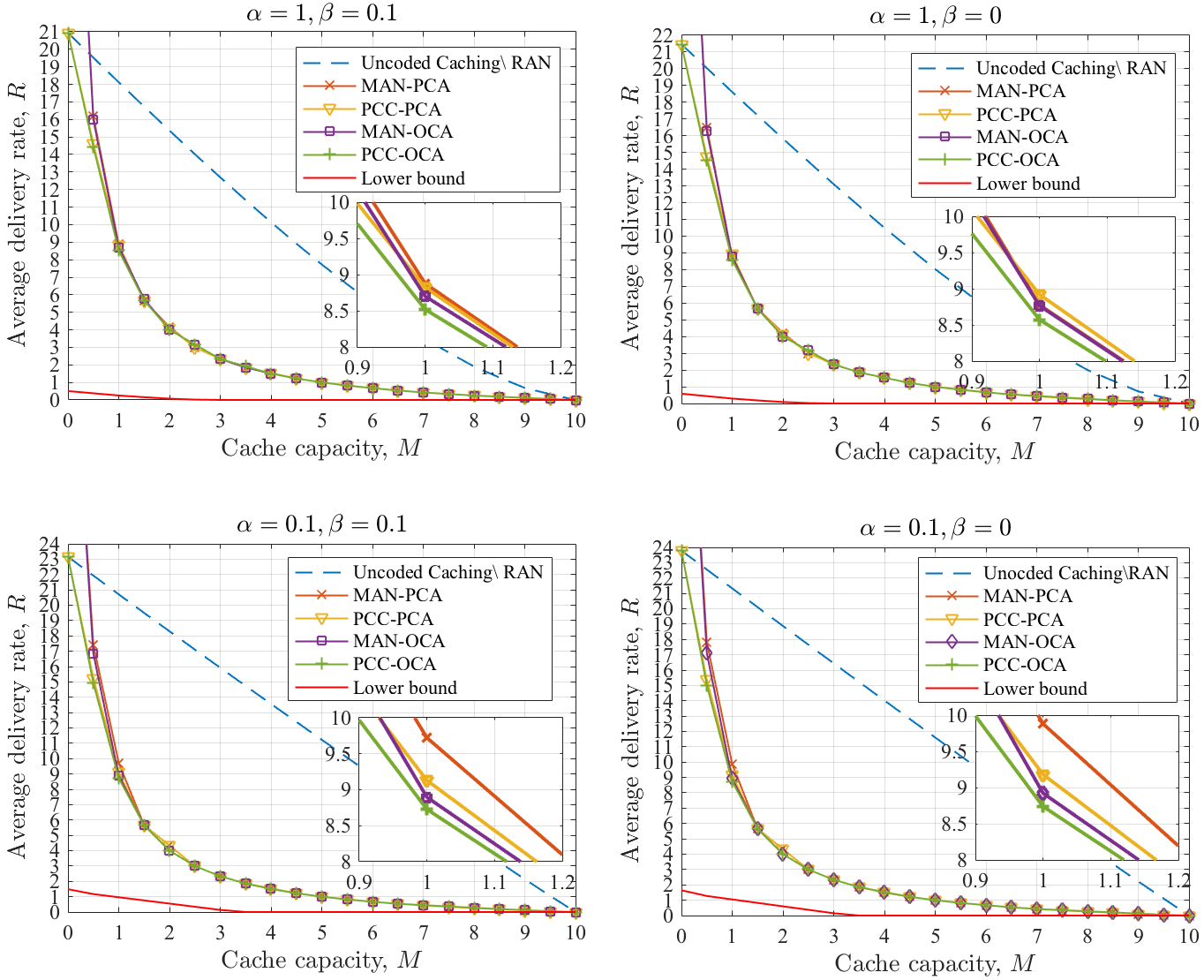}
\caption{Comparison between PCC, MAN, uncoded caching and the lower bound given different values of $\alpha$ and $\beta$.}
\label{comp1}
\end{figure}

\begin{figure}
\centering
%width=1.0\linewidth
\includegraphics[scale=0.75]{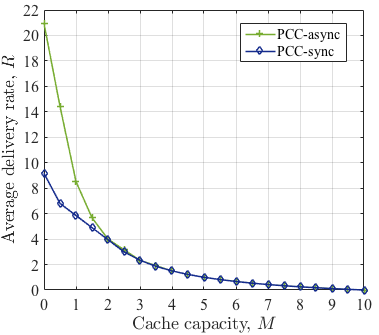}
%trim={left bottom right top}
\caption{Comparison between the asynchronous and synchronous demand arrival scenarios, $\alpha=1$ and $\beta=0.1$.}
\label{comp2}
\end{figure}

We set $A_{\mathrm{max}}=15$, and $P_A(15)=1$; that is, exactly $15$ new demands arrive at each time slot. It is verified in Fig.~\ref{comp1} that, the RAN and Uncoded schemes have the same performance. We also observe from Fig.~\ref{comp1} that both PCC and MAN significantly reduce the average delivery rate compared to Uncoded and RAN, and the improvement increases with the cache capacity. We can see that the PCC scheme notably outperforms MAN scheme when the cache capacity is small as the PCC scheme is more efficient in delivering the bits that either have not been cached by any user, or have been cached exclusively by one user. An interesting observation is that, both PCC and MAN achieve almost the same performance with either of the two cache allocation schemes, PCA and OCA. This implies that caching as many of the most popular chunks as possible can be sufficient to fully exploit the cache capacities. However, slight improvement of PCA over OCA can be observed in the zoomed-in subfigures in Fig. \ref{comp1}. We can also observe that a larger $\alpha$ results in a smaller average delivery rate since the users tend to request the most popular files, and caching these files is more efficient in terms of the average delivery rate. In contrast, a smaller $\beta$ increases the average delivery rate since users tend to continue watching their requests, which increases the overall demand. We also note that the gap between the lower bound and the achievable delivery rate remains significant, which calls for more research. 

We evaluate the effect of the asynchronous arrival of demands by considering two scenarios: in the first scenario, $15$ new users arrive at each time slot as in Fig. \ref{comp1}; while in the second one, $45$ new users arrive at every three time slots, while the demands are asynchronous in the first scenario, they are synchronized in the second as all the active users watch the same chunk. The average delivery rates achieved by PCC-OCA are shown in Fig. \ref{comp2} for the two scenarios, labeled as \textit{PCC-async} and \textit{PCC-sync}. We see that PCC-sync has remarkably lower average delivery rates than PCC-async when the cache capacity is small, since there are less distinct demands in each time slot. However, as the cache capacity increases, the effect of distinct demands is compensated since coded delivery can create multicasting opportunities by exploiting the cached contents. Hence, we can conclude from Fig. \ref{comp2} that larger cache capacities are needed to observe the benefits of coded delivery in the more realistic setting of asynchronous user demands.

\begin{figure}[!t]
\centering
\includegraphics[width=.7\linewidth]{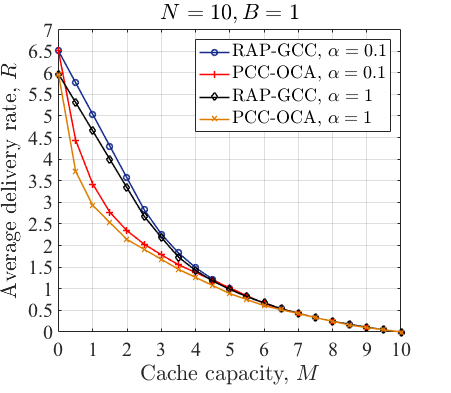}
\caption{Comparison between PCC with OCA and RAP-GCC with $\alpha=0.1$ and $\alpha=1$.}
\label{rapgcc}
\end{figure}

In Fig. \ref{rapgcc}, we compare the performance of PCC with the RAP-GCC scheme in \cite{JiArXivNonuniform}, which, to the best of our knowledge, is the only result in the literature on the average delivery rate considering heterogeneous file popularities. We set $B=1$, such that the partial caching problem studied in this paper reduces to the one in \cite{JiArXivNonuniform}. For fair comparison, we optimize the cache content distribution for the RAP-GCC scheme as well. It is notable in Fig. \ref{rapgcc} that PCC outperforms RAP-GCC, and as $\alpha$ becomes larger, i.e, the popularity distribution of the files becomes more skewed, the gap between the two schemes increases slightly.

\section{Conclusions}\label{SecConc}

We have studied content caching and coded delivery in a more realistic system model, allowing asynchronous demand arrivals, and taking into account the audience retention rates of the video files; that is, we allow the users to dynamically join the system, place a request, consume a random portion of the request, and leave the system at a random time. We assume that each video file in the library consists of a number of chunks of the same size, and the audience retention rate is modeled as the heterogeneous popularity of the chunks of each file. We proposed a coded caching scheme that allocates users' cache capacities to different chunks, depending on their popularities. We then evaluated the average delivery rate over all possible demand combinations. We proposed two different methods for cache allocation, namely, the numerically optimized cache allocation scheme OCA, and a low complexity popularity-based cache allocation scheme PCA. The numerical results showed a significant improvement with the proposed scheme over uncoded caching in terms of the average delivery rate, or the extension of other known delivery methods to the asynchronous scenario. We have also derived an information theoretic lower bound on the average delivery rate.

\appendices
\section{Proof of Theorem \ref{theorem:rate0}}
Since each user requesting chunk $W_{ij}$ already has $q_{ij}F/B$ bits of it cached, according to \cite[Appendix A]{MaddahAliDecentralized}, at most $(1-q_{ij})F/B+o(F/B)$ bits are necessary to enable all the users requesting $W_{ij}$ to decode it, for $i \in [N]$ and $j \in [B]$. The probability that chunk $W_{ij}$ is requested by at least one user at time slot $t$ is given by:
\begin{equation}\label{eqij}
\mathrm{Pr}\{W_{ij} \in \mathcal{D}_t\}=\sum\nolimits_{k \in \mathcal{A}}\mathrm{Pr}\{K_j=k\} \Big(1-\left(1-\tilde{p}_{ij}\right)^k\Big). 
\end{equation}
By summing over $i\in [N]$ and $j \in [B]$, and ignoring the $o(F/B)$ term, we complete the proof:
\begin{equation}
R_{\rm{RAN}}(P_A, \mathbf{p}, \mathbf{P}, \mathbf{Q})= \sum\nolimits_{j=1}^B \sum\nolimits_{i=1}^N\sum\nolimits_{k \in \mathcal{A}}\mathrm{Pr}\{K_j=k\} \Big(1-\left(1-\tilde{p}_{ij}\right)^K\Big)(1-q_{ij}).
\end{equation}

\section{Proof of Theorem \ref{theorem:rate1}}
Recall that $K_j$ is the number of users demanding the $j$-th chunks of their requested files at time slot $t$, and these $K_j$ users are indexed with $\left[K'_{j-1}+1: K'_j\right]$, where $K'_j\triangleq \sum\nolimits_{s=1}^{j}K_s$ and $K'_0\triangleq 0$, for $j\in [B]$. Similar to the proof in \cite[Appendix A]{JiArXivNonuniform}, the average number of bits (normalized by $F/B$) sent by the MAN scheme over all possible demand combinations is given by 
\begin{subequations}\label{RMAN}
\begin{align}
&R_{\rm{MAN}}^t(K_{[B]})=\mathbb{E}\left[\sum\nolimits_{z=0}^{K^{(t)}}\sum\nolimits_{\substack{\mathcal{P} \subset [K^{(t)}],|\mathcal{P}|=z}} \max\limits_{k\in \mathcal{P}}\Big| W^t_{d_{k, t}, \mathcal{P}\setminus\{k\}}\Big |\Bigg| C\right]\label{ra2:1}\\
&=\sum\nolimits_{\mathbf{d}_{t} \in [N]^{K_1}\times [N]^{K_2} \times \cdots \times [N]^{K_B}} \left(\prod\nolimits_{j=1}^{B}\prod\nolimits_{k=1}^{K_j} \tilde{p}_{d_{K'_{j-1}+k, t}}\right)\nonumber\\
&\left(\sum\limits_{l_{[B]}\in [0:K_1]\times \cdots \times [0:K_B]}\sum\limits_{\substack{\mathcal{P}_1 \subset [K'_{0}+1: K'_{1}]: \\|\mathcal{P}_1|=l_1}}\sum\limits_{\substack{\mathcal{P}_2 \subset [K'_{1}+1: K'_{2}]: \\|\mathcal{P}_2|=l_2}}\cdots\sum\limits_{\substack{\mathcal{P}_B \subset [K'_{B-1}+1: K'_{B}]: \\|\mathcal{P}_B|=l_B}}\max\limits_{\substack{k\in \mathcal{P}:\\\mathcal{P}=\bigcup\limits_{s=1}^{B}\mathcal{P}_s}}\Big|W^t_{d_{k, t}, \mathcal{P}\setminus\{k\}}\Big |\right)\label{ra2:2}\\
&=\sum\nolimits_{l_{[B]}\in [0:K_1]\times \cdots \times [0:K_B]}\left(\prod\nolimits_{j=1}^B\binom{K_j}{l_j}\right)\sum\nolimits_{(d_{1, t}, ..., d_{l_1, t})\in [N]^{l_1}}\cdots \sum\nolimits_{(d_{K'_{B-1}+1, t}, ..., d_{K'_{B-1}+l_B, t})\in [N]^{l_B}}\nonumber\\
&\left(\prod_{j=1}^B\prod_{k=1}^{l_j} \tilde{p}_{d_{K'_{j-1}+k, t}}\right)\sum\limits_{\substack{k \in \mathcal{P}:\\\mathcal{P}=\bigcup\limits_{s=1}^{B}\mathcal{P}_s, \mathcal{P}_s=[K'_{s-1}+1:K'_{s-1}+l_s]}}\frac{\mathbbm{1}\left\{ k=\argmax\limits_{h \in \mathcal{P}}\Big |W^t_{d_{h, t}, \mathcal{P}\setminus\{h\}}\Big |\right\}}{\sum\limits_{h \in \mathcal{P}}\mathbbm{1}\left\{ \Big |W^t_{d_{h, t}, \mathcal{P}\setminus\{h\}}\Big |=\Big |W^t_{d_{k, t}, \mathcal{P}\setminus\{k\}}\Big |\right\}}\cdot\Big |W^t_{d_{k, t}, \mathcal{P}\setminus\{k\}}\Big |\label{ra2:3}\\
&=\sum\nolimits_{l_{[B]}\in [0:K_1]\times \cdots \times [0:K_B]}\left(\prod\nolimits_{j=1}^B\binom{K_j}{l_j}\right)\sum\nolimits_{i=1}^{N}\sum\nolimits_{j=1}^{B}\nonumber\\
&~~~~\frac{\mathrm{Pr}\left(\max\limits_{W_{fh} \in \bigcup\limits_{s=1}^{B}\{W_{d_{K'_{s-1}+1, t}}, ..., W_{d_{K'_{s-1}+l_s, t}}\}}\left| W^t_{fh,[ \sum\limits_{s=1}^Bl_s-1]}\right|=\left| W^t_{ij,[ \sum\limits_{s=1}^Bl_s-1]}\right|\right)}{\sum\nolimits_{f=1}^{N}\sum\nolimits_{h=1}^{B}\mathbbm{1}\left\{ \Big |W^t_{fh,[ \sum\nolimits_{s=1}^Bl_s-1]}\Big |=\Big |W^t_{ij,[ \sum\nolimits_{s=1}^Bl_s-1]}\Big |\right\}}\left| W^t_{ij,[ \sum\limits_{s=1}^Bl_s-1]}\right|\label{ra2:4}
\end{align}
\end{subequations}
where $C$ is the realization of cache contents for a fixed cache content distribution $\mathbf{Q}$; \eqref{ra2:2} is derived by finding the expectation over all possible demand realizations $\mathbf{d}_{t}$ given $K_{1}, ..., K_{B}$. The probability of any demand combination $\mathbf{d}_{t} \in [N]^{K_1}\times [N]^{K_2} \times \cdots \times [N]^{K_B}$ is thus $\prod_{j=1}^{B}\prod_{k=1}^{K_j} \tilde{p}_{d_{K'_{j-1}+k, t}}$.  \eqref{ra2:2} also specifies the number of users in $\mathcal{P}$ requesting different chunks, i.e., $l_1, ..., l_B$, where $l_j$ is the number of users in $\mathcal{P}$ demanding the $j$-th chunks, for $j\in [B]$. Notice that 
\begin{equation}\label{kk}
\max\nolimits_{\substack{k\in \mathcal{P}:\mathcal{P}=\bigcup\nolimits_{s=1}^{B}\mathcal{P}_s}}\Big|W^t_{d_{k, t}, \mathcal{P}\setminus\{k\}}\Big |
\end{equation}
depends only on the demands of users in $\mathcal{P}$. \eqref{ra2:3} follows by first changing the order of the summation, which is to choose a set of $\sum\nolimits_{s=1}^B l_s$ users first, among which $l_j$ users request their $j$-th chunks, for $j \in [B]$, and then take the expectation of the number of bits sent to this set of users over all possible demand combinations. Note that, due to the symmetry across users, for a given $l_j$, $j \in [B]$, any $l_j$ users among $K_j$ can be considered. Henceforth, for any $(l_1, ..., l_B) \in [0:K_1]\times \cdots \times [0:K_B]$, \eqref{ra2:3} only considers the first $l_j$ users, i.e., users $K'_{j-1}+1, ..., K'_{j-1}+l_j$, among the $K_j$ users demanding the $j$-th chunks, for $j \in [B]$, without any loss of accuracy. Writing the expectation with regards to each chunk yields \eqref{ra2:4}, where we note that if $\sum\nolimits_{s=1}^Bl_s=0$, $\left| W^t_{ij,\left[ \sum\nolimits_{s=1}^Bl_s-1\right]}\right|=0$, $\forall i\in [N], \forall j\in [B]$. We emphasize that the probability 
\[\mathrm{Pr}\left(\max\nolimits_{W_{fh} \in \bigcup\nolimits_{s=1}^{B}\{W_{d_{K'_{s-1}+1, t}}, ..., W_{d_{K'_{s-1}+l_s, t}}\}}\left| W^t_{fh,[ \sum\nolimits_{s=1}^Bl_s-1]}\right|=\left| W^t_{ij,[ \sum\nolimits_{s=1}^Bl_s-1]}\right|\right)\]
is taken over all the possible realizations of $\bigcup\nolimits_{s=1}^{B}\{d_{K'_{s-1}+1, t}, ..., d_{K'_{s-1}+l_s, t}\}$, which is distributed according to $\mathbf{p}$ and $\mathbf{P}$, and is equivalent to $\rho_{ij, (K_{[B]},l_{[B]})}$ as defined in \eqref{defi:rho}. We also remark that, in \eqref{ra2:4}, the expected number of bits sent to any subset of users specified by $(l_{[B]})$ is calculated with respect to the first $\sum\nolimits_{s=1}^Bl_s-1$ users, i.e., users $\left[ \sum\nolimits_{s=1}^Bl_s-1\right]$. That is because the number of bits of each chunk cached exclusively by any subset of $\sum\nolimits_{s=1}^Bl_s-1$ users among given $\sum\nolimits_{s=1}^BK_s$ users is almost identical according to the law of large number.  It can be concluded from \eqref{ra2:4} that the value of $R_{\rm{MAN}}^t(K_{[B]})$ is irrelevant to $t$, given $K_{[B]}$. Thus, we simply use $R_{\rm{MAN}}(K_{[B]})$ in the sequel. 

Given the cache content distribution $\mathbf{Q}$ and $K_{[B]}$, the size of $W^t_{ij,\left[ \sum\nolimits_{s=1}^Bl_s-1\right]}$ is  
\begin{subequations}
\begin{align}
\Big| W^t_{ij,\left[ \sum\nolimits_{s=1}^Bl_s-1\right]}\Big|&=(q_{ij})^{\sum\nolimits_{s=1}^Bl_s-1}(1-q_{ij})^{\sum\nolimits_{s=1}^BK_s-\sum\nolimits_{s=1}^Bl_s+1}+o(F/B)\\
&=g_{ij, (\sum\nolimits_{s=1}^{B}K_s, \sum\nolimits_{s=1}^{B}l_s-1)}+o(F/B).
\end{align}
\end{subequations}
Ignoring the term $o(F/B)$ and substituting $\Big| W^t_{ij,[ \sum\nolimits_{s=1}^Bl_s-1]}\Big|$ and $\Big| W_{fh,[ \sum\nolimits_{s=1}^Bl_s-1]}\Big|$ in \eqref{ra2:4} yields  
\begin{equation}\label{MANrate_1}
R_{\rm{MAN}}(K_{[B]}) = \sum\limits_{l_{[B]}\in [0:K_1]\times \cdots \times [0:K_B]}\left(\prod\limits_{j=1}^B\binom{K_j}{l_j}\right)\sum\limits_{i=1}^{N}\sum\limits_{j=1}^{B}\rho'_{ij, (K_{[B]},~l_{[B]})}g_{ij, (\sum\limits_{s=1}^{B}K_s, \sum\limits_{s=1}^{B}l_s-1)},
\end{equation}
where $\rho'_{ij, ((K_{[B]}),~(l_{[B]}))}$ is defined in \eqref{defi:rhoprime}. Taking the expectation over all possible realizations of $(K_{[B]})$, we obtain the average delivery rate of Algorithm~\ref{deliveryscheme1} given as follows: 
\begin{subequations}\label{MANrate}
\begin{align}
R_{\rm{MAN}}&=\sum\nolimits_{k_{[B]}\in \mathcal{A}^B}\left(\prod\nolimits_{s=1}^B \mathrm{Pr}\{K_s=k_s\}\right)R_{\rm{MAN}}(k_{[B]})\\
&=\sum\nolimits_{k_{[B]}\in \mathcal{A}^B}\left(\prod\nolimits_{s=1}^B \mathrm{Pr}\{K_s=k_s\}\right)\sum\nolimits_{(l_{[B]})\in [0:k_1]\times \cdots \times [0:k_B]}\left(\prod\nolimits_{j=1}^B\binom{k_j}{l_j}\right)\cdot\nonumber\\
&~~~\sum\nolimits_{i=1}^{N}\sum\nolimits_{j=1}^{B} \rho'_{ij, (k_{[B]},~l_{[B]})}g_{ij, (\sum\nolimits_{s=1}^{B}k_s, \sum\nolimits_{s=1}^{B}l_s-1)},
\end{align}
\end{subequations}
which completes the proof. 

\section{Proof of Theorem \ref{theorem:rate}}

We first prove $\Delta \varphi_1(P_A, \mathbf{p}, \mathbf{P}, \mathbf{Q})$ given in \eqref{deltavarphi1}, which is the difference between the number of bits sent by PART 1 of Algorithm~\ref{deliveryscheme} and those sent by the MAN scheme for $z=1$, both averaged over all the demand combinations. We then derive $\Delta \varphi_2(P_A, \mathbf{p}, \mathbf{P}, \mathbf{Q})$ given in \eqref{deltavarphi2}, which is the difference between the number of bits sent by PART 2 of Algorithm~\ref{deliveryscheme} and those delivered by the MAN scheme for $z=2$, both averaged over all the demand realizations.

In PART 1 of Algorithm~\ref{deliveryscheme}, the server sends the missing bits which are not cached by any user in $[K^{(t)}]$. The expected number of bits of chunk $W_{ij}$ that are not cached by any user in $[K^{(t)}]$ is given by 
\begin{equation}\label{ExpectedNoWhere}
F/B(1-q_{ij})^{K^{(t)}}+o(F/B).
\end{equation}
Recall that $K_j$ denotes the number of users demanding the $j$-th chunks, $j \in [B]$; i.e., $K^{(t)}=\sum\nolimits_{s=1}^BK_s$. The probability that chunk $W_{ij}$ is requested by at least one user at the beginning of time slot $t$ is given by \eqref{eqij}. By summing over $i \in [N]$ and $j \in [B]$, ignoring $o(F/B)$ term, and taking the expectation over all realizations of $K_{[B]}$, we obtain the average number of bits delivered in PART 1 of Algorithm \ref{deliveryscheme} as:
\begin{equation}\label{varphi_1}
\varphi_1=\sum\nolimits_{k_{[B]}\in \mathcal{A}^B}\left(\prod\nolimits_{s=1}^B \mathrm{Pr}\{K_s=k_s\}\right)\sum\nolimits_{j=1}^B \sum\nolimits_{i=1}^N\Big(1-(1-\tilde{p}_{ij})^{k_{j}}\Big)g_{ij, (\sum\nolimits_{s=1}^B k_s, 0)}.
\end{equation}
Next, we derive the average number of bits sent by Algorithm \ref{deliveryscheme1} for $z=1$, denoted by $\overline{\varphi}_1$. Following the similar procedure of the proof of \eqref{MANrate}, we have 
\begin{equation}\label{varphi_1https://www.sharelatex.com/project/59427c2a1c8136581e1d20d1_MAN}
\overline{\varphi}_1=\sum\nolimits_{k_{[B]}\in \mathcal{A}^B}\left(\prod\nolimits_{s=1}^B \mathrm{Pr}\{K_s=k_s\}\right)\sum\nolimits_{j=1}^Bk_j\sum\nolimits_{i=1}^N\tilde{p}_{ij}g_{ij, (\sum\nolimits_{s=1}^B k_s, 0)}.
\end{equation}
Thus, we have $\Delta \varphi_1(P_A, \mathbf{p}, \mathbf{P}, \mathbf{Q})= \overline{\varphi}_1 - \varphi_1$, which proves \eqref{deltavarphi1}.

Recall that the probability that chunk $W_{ij}$ is requested by at least one user at the beginning of time slot $t$ is given by \eqref{eqij}. Hence, the average number of bits sent by PART 2.2 of Algorithm \ref{deliveryscheme} is given by          
\begin{equation}\label{varphi_3_1}
\varphi_{2}=\sum\nolimits_{k_{[B]}\in \mathcal{A}^B}\left(\prod\nolimits_{s=1}^B \mathrm{Pr}\{K_s=k_s\}\right)\sum\limits_{j=1}^B \sum\limits_{i=1}^N(\sum\limits_{s=1}^B k_s-1) \left(1-\left(1-\tilde{p}_{ij}\right)^{k_{j}}\right)g_{ij, (\sum\nolimits_{s=1}^B k_s, 1)}.
\end{equation}
Following similar steps to the proof of \eqref{MANrate}, the number of bits sent by Algorithm \ref{deliveryscheme1} for $z=2$ (or PART 2.1 of Algorithm \ref{deliveryscheme}) is given by 
\begin{align}\label{varphi_3}
&\overline{\varphi}_{2}=\sum\nolimits_{k_{[B]}\in \mathcal{A}^B}\left(\prod\nolimits_{s=1}^B \mathrm{Pr}\{K_s=k_s\}\right)\left(\sum\nolimits_{j=1}^B\binom{k_j}{2}\sum\nolimits_{i=1}^N\rho'_{ij, (k_{[B]},~(0, ..., l_j=2, ..., 0))} g_{ij, (\sum\nolimits_{s=1}^B K_s, 1)}\right.\nonumber\\
&\; \left.+\sum\nolimits_{j_1=1}^B k_{j_1}\sum\nolimits_{j_2=j_1+1}^Bk_{j_2}\sum\nolimits_{i=1}^N\sum\nolimits_{j=1}^B\rho'_{ij, (k_{[B]},~(0, ..., l_{j_1}=1, ..., l_{j_2}=1, ..., 0))}g_{ij, (\sum\nolimits_{s=1}^Bk_s, 1)}\right).
\end{align}
Thus, we have $\Delta \varphi_2(P_A, \mathbf{p}, \mathbf{P}, \mathbf{Q})=\max\{ \overline{\varphi}_2 - \varphi_2, 0\}$, which proves \eqref{deltavarphi2}, and completes the proof of Theorem \ref{theorem:rate}.

\section{Proof of Theorem \ref{lowerbound}}

To prove Theorem \ref{lowerbound}, we first derive a lower bound on the optimal rate of any time slot $t$ given the number of users watching different chunks, i.e., $K_{1}, ..., K_{B}$, averaged over all possible demand combinations for these users, denoted by $R_{\rm{opt}}(K_{[B]}, M)$. We have
\small
\begin{equation}
R_{\rm{opt}}(K_{[B]}, M) \buildrel \Delta \over =
\inf\left\{\mathbb{E}\left[ R_{\mathbf{d}_{t}}(M)\Big|K_{[B]}, C\right] \right\},
\end{equation}
\normalsize
where the infimum is taken over all the achievable schemes, and the expectation is taken over all possible demand configurations $\mathbf{d}_t$, distributed according to $\mathbf{p}$ and $\mathbf{P}$, given $K_{[B]}$. We recall that these users are re-indexed such that users $K'_{j-1}+1, ..., K'_{j}$ demand the $j$-th chunks at current time slot, for $j \in [B]$. %Since $\mathbf{p}$ and $\mathbf{P}$ are time invariant, the value of $R_{opt}^t(a_{t-1}, ..., a_{t-B})$ only depends on the values of $a_{t-1}, ..., a_{t-B}$. Hence, we use $R_{opt}(a_{1}, ..., a_{B})$ instead of $R_{opt}^t(a_{t-1}, ..., a_{t-B})$.

Inspired by \cite[Appendix C]{JiArXivNonuniform}, in order to lower bound $R_{\rm{opt}}(K_{[B]}, M)$, we consider the following genie-aided system: we recall that $r_{1j}, ..., r_{Nj}$ is an ordered permutation of $\{\tilde{p}_{1j}, ..., \tilde{p}_{Nj}\}$, such that $r_{1j}\geq \cdots \geq r_{Nj}$, $\forall j \in [B]$. For $j \in [B]$, fix $n_j \in [N]$. As aforementioned, considering user $k$ demanding a $j$-th chunk, i.e., $k\in [K'_{j-1}+1:K'_{j}]$, if the requested $j$-th chunk has a normalized popularity  lower than $r_{n_jj}$, i.e.,  $\tilde{p}_{d_{k, t}}<r_{n_jj}$, it is served by a genie at no transmission cost; otherwise, i.e., if $\tilde{p}_{d_{k, t}}\geq r_{n_jj}$, it is served by a genie at no transmission cost with probability $1-r_{n_jj}/\tilde{p}_{d_{k, t}}$; that is, the server has to transmit the required $j$-th chunk to this user through the shared link with probability $r_{n_jj}/\tilde{p}_{d_{k, t}}$. Thus, each user demanding a $j$-th chunk requires service from the server, i.e., not from the genie, with probability $n_jr_{n_jj}$. This immediately implies that the total number of users who are demanding the $j$-th chunks, and served by the server during time slot $t$, denoted by $V_j$, follows a Binomial distribution $\mathrm{Binomial}(K_{j}, n_jr_{n_jj})$, i.e., $\mathrm{V}_j\sim\mathrm{Binomial}(K_{j}, n_jr_{n_jj})$.

%Since the distribution of $a_{t-j}$ is identical for any $t-j\leq 1$, $t\in \mathbb{N}$, $j\in[B]$, we use $a_{j}$ instead of $a_{t-j}$ for simplicity. Since $\mathbf{p}$ and $\mathbf{P}$ are identical across users and invariant over time, the value of $R_{\rm{opt}}^t(a_{t-B:t-1})$ depends only on $\{a_{t-B:t-1}\}$. Thus, for simplicity, we use $R_{\rm{opt}}(K_{[B]})$ instead of $R_{\rm{opt}}^t(a_{t-B:t-1})$ in the sequel. 

We denote the optimal rate of the above genie-aided system by $R_{\rm{genie\_opt}}(K_{[B]}, n_{[B]}, M)$. For any $n_{[B]}\in [N]^B$, it  provides a lower bound on the optimal rate of the original system, i.e., $R_{\rm{opt}}(K_{[B]}, M)$, since a subset of users are served by the genie. Note that, for the genie-aided system, the demands of the $j$-th chunks that are served by the server instead of the genie are independent and uniformly distributed over all the $j$-th chunks with a normalized popularity no less than $r_{n_jj}$, i.e., $\{W_{ij}: \tilde{p}_{ij}\geq r_{n_jj}, i\in [N] \}$, the cardinality of which is $n_j$ according to the definition of $r_{ij}$, $\forall j \in [B]$. That is, for $ k\in [K'_{j-1}+1: K'_{j}]$, 
\begin{align}
\mathrm{Pr}(d_{k, t}=ij| \mbox{the $k$-th user requires service from server})\triangleq
\begin{cases}
1/n_j, \quad &\mbox{if $\tilde{p}_{ij} \geq r_{n_jj}$};\\
0,\quad &\mbox{if $\tilde{p}_{ij}< r_{n_jj}$},
\end{cases}
\end{align}
$\forall i\in [N], j\in [B]$. Let $R_{\rm{opt\_unif}}(v_{[B]}, n_{[B]}, M)$ denote the optimal rate of a system including $\sum\nolimits_{j=1}^{B} v_j$ users, each equipped with a cache of size $MF$ bits, where each user in a unique subset of $v_j$ users among them independently demands one chunk from a subset of $n_j$ $j$-th chunks with uniform popularity distribution, for $j \in [B]$. It follows that
\begin{subequations}\label{unifrate}
\begin{align}
&R_{\rm{genie\_opt}}(K_{[B]}, n_{[B]}, M)\geq \mathbb{E}\left(R_{\rm{opt\_unif}}(\mathrm{V}_{[B]}, n_{[B]}, M)\right)\label{unifrate1}\\
&\quad \qquad \qquad \qquad =\sum\nolimits_{V_{[B]} \in [K_1]\times\cdots \times [K_B]} \left(\prod\nolimits_{j=1}^B \mathrm{Pr}(\mathrm{V}_j=V_j)\right)R_{\rm{opt\_unif}}(V_{[B]}, n_{[B]}, M)\label{unifrate2}\\
&\quad \qquad \qquad \qquad \geq\sum\nolimits_{V_1=v_1}^{K_1}\cdots\sum\nolimits_{V_B=v_B}^{K_B} \left(\prod\nolimits_{j=1}^B \mathrm{Pr}(\mathrm{V}_j=V_j)\right)R_{\rm{opt\_unif}}(V_{[B]}, n_{[B]}, M)\label{unifrate3}\\
&\quad \qquad \qquad \qquad \geq \left(\prod\nolimits_{j=1}^B \mathrm{Pr}(\mathrm{V}_j\geq v_j)\right) R_{\rm{opt\_unif}}(v_{[B]}, n_{[B]}, M),\label{unifrate4}
\end{align}
\end{subequations}
where the expectation in \eqref{unifrate1} is taken over all the values of $\mathrm{V}_{[B]}$, which yields \eqref{unifrate2}; \eqref{unifrate3} is derived by deleting some non-negative terms; \eqref{unifrate4} is due to the fact that the optimal rate is non-decreasing with the number of users. 

In the following, we lower bound $R_{\rm{opt\_unif}}(v_{[B]}, n_{[B]}, M)$ by applying \cite[Lemma 4]{NiesenNonuniform}. 
\begin{Lemma}
$R_{\rm{opt\_unif}}(v_{[B]}, n_{[B]}, M)$ defined above should satisfy
\begin{equation}\label{lemma1}
R_{\rm{opt\_unif}}(v_{[B]}, n_{[B]}, M)\geq \prod\nolimits_{j=1}^B \mathrm{Pr}(\mathrm{Z}_j\geq z_j) R_{\rm{opt}}(z_{[B]}, n_{[B]}, M),
\end{equation}
for any $z_{[B]}$, such that $z_j \in [\min \{v_j, n_j\}]$, for $j \in [B]$, where $\mathrm{Z}_j$ is a random variable indicating the number of distinct $j$th chunks requested by $v_j$ users from a library of $n_j$ $j$th chunks with a uniform popularity distribution. Furthermore, $R_{\rm{opt}}(z_{[B]}, n_{[B]}, M)$ is the expected rate of the optimal scheme with $z_j$ distinct demands of the $j$th chunks selected uniformly at random from $n_j$ $j$th chunks, for $j \in [B]$. 
\end{Lemma}
Below, we derive a lower bound on $R_{\rm{opt}}(z_{[B]}, n_{[B]}, M)$ following the cut-set technique. Since the delivery rate is non-decreasing with the number of users, we restrict to a subset of users $\mathcal{U}$ consisting of $\sum\nolimits_{j=1}^{B} z_j$ users, where a distinct subset of $z_j$ users among them request $z_j$ distinct chunks from a subset of $n_j$ $j$th chunks with uniform popularity, for $j \in [B]$. We note that there exist $\prod\nolimits_{j=1}^B \binom{n_j}{z_j}z_j!$ demand combinations of these users, each of identical probability due to the uniform distribution of chunks. We group these demand combinations into $G_{tol}\triangleq\frac{\prod\nolimits_{j=1}^B \binom{n_j}{z_j}z_j!}{\min\nolimits_{j\in [B]}\lfloor n_j/z_j \rfloor}$ disjoint groups, denoted by $\mathfrak{G}_1, ..., \mathfrak{G}_{G_{tol}}$, such that each group consists of $\min\nolimits_{j\in [B]}\lfloor n_j/z_j \rfloor$ disjoint demand combinations. Consider one such group $\mathfrak{G}_g$, $g\in [G_{tol}]$. For a demand combination in this group and a corresponding message over the shared link, say $X^g_1$, $X^g_1$ and $\{Z_K^{(t)}\mid k\in \mathcal{U}\}$ allow the reconstruction of a subset of $\sum\nolimits_{j=1}^B z_j$ chunks; similarly, for another demand combination in this group and a corresponding input to the shared link, say $X^g_2$, $X^g_2$ and $\{Z_K^{(t)}\mid k\in \mathcal{U}\}$ allow the reconstruction of another disjoint subset of $\sum\nolimits_{j=1}^B z_j$ chunks; and so on so forth. Hence, with $X^g_1$, ..., $X^g_{\min\nolimits_{j\in [B]}\lfloor n_j/z_j \rfloor}$ and $\{Z^n_k\mid k\in \mathcal{U}\}$, each user $k \in \mathcal{U}$ can reconstruct a distinct set of $\min\nolimits_{j\in [B]}\lfloor n_j/z_j \rfloor$ chunks. By considering a cut separating $X^g_1$, ..., $X^g_{\min\nolimits_{j\in [B]}\lfloor n_j/z_j \rfloor}$ and $\{Z^n_k\mid k\in \mathcal{U}\}$ from the corresponding users, we have \cite[Theorem 14.10.1]{CoverInformation}
\begin{align}\label{theo1:proof1}
 \sum\nolimits_{i=1}^{\min\nolimits_{j\in [B]}\lfloor n_j/z_j \rfloor} |X^g_i|+\sum\nolimits_{k\in \mathcal{U}} |Z_K^{(t)}| \geq  \sum\nolimits_{j=1}^{B}z_j \min\nolimits_{j\in [B]}\lfloor n_j/z_j \rfloor.
\end{align}
We have 
\begin{align}
R_{\rm{opt}}(z_{[B]}, n_{[B]}, M) = \inf\left\{\frac{1}{G_{tol}}\sum\nolimits_{g=1}^{G_{tol}} \sum\nolimits_{i=1}^{\min\nolimits_{j\in [B]}\lfloor n_j/z_j \rfloor} \frac{|X^g_i|}{\min\nolimits_{j\in [B]}\lfloor n_j/z_j\rfloor}\right\},    
\end{align}
where the infimum is is taken over all the achievable schemes. We also have the cache capacity constraints $MB \geq |Z_K^{(t)}|$ (normalized by $F/B$), for $k \in \mathcal{U}$. Plugging these into \eqref{theo1:proof1}, we obtain
\begin{align}\label{roptznm}
R_{\rm{opt}}(z_{[B]}, n_{[B]}, M)\geq \sum\nolimits_{j=1}^B z_j \left(1-\frac{MB}{\min\nolimits_{j\in [B]}\lfloor n_j/z_j \rfloor}\right).
\end{align}

Next, we note that both $\mathrm{V}_j$ and $\mathrm{Z}_j$ are random variables expressed as self-bounding functions of random vectors (see \cite[Definition 3]{JiArXivNonuniform}). We apply a concentration property of these random variables (see \cite[Lemma 4]{JiArXivNonuniform}) to lower bound probabilities $\mathrm{Pr}(\mathrm{V}_j\geq v_j)$ and $\mathrm{Pr}(\mathrm{Z}_j\geq z_j)$, and find the range of $v_j$ and $z_j$, for $j \in [B]$. According to \cite[Lemma 4]{JiArXivNonuniform}, we can write 
\begin{equation}\label{proof_theorem2_pk}
\mathrm{Pr}\left(\mathrm{V}_j\geq \mathbb{E}\left[\mathrm{V}_j\right]-\mu\right)\geq 1- \exp\left(-\frac{\mu^2}{2\mathbb{E}\left[\mathrm{V}_j\right]}\right),
\end{equation}
with $0<\mu\leq \mathbb{E}\left[\mathrm{V}_j\right]$. We have $\mathbb{E}\left[\mathrm{V}_j\right]= K_jn_jr_{n_jj}$ as $\mathrm{V}_j\sim\mathrm{Binomial}(K_j, n_jr_{n_jj})$. Letting $\mu= \mathbb{E}\left[\mathrm{V}_j\right]-v_j$, we obtain
\begin{equation}\label{proof_theorem2_pk2}
\mathrm{Pr}\left(\mathrm{V}_j\geq v_j\right)\geq  1-\exp\left(-\frac{(K_jn_jr_{n_jj}-v_j)^2}{2K_jn_jr_{n_jj}}\right)\triangleq f'_j(K_j, n_j, v_j),
\end{equation}
where $0<v_j\leq K_jn_jr_{n_jj}$, for $j\in [B]$. Similarly, we have 
\begin{equation}\label{proof_theorem2_pz2}
\mathrm{Pr}\left(\mathrm{Z}_j\geq z_j\right)\geq  1-\exp\left(-\frac{(f(n_j, v_j)-z_j)^2}{2f(n_j, v_j)}\right)\triangleq f''_j(n_j, v_j, z_j),
\end{equation}
where $\mathbb{E}\left[\mathrm{Z}_j\right]= n_j\left(1-\left(1-1/n_j\right)^{v_j}\right)\triangleq f(n_j, v_j)$, for $0<z_j\leq f(n_j, v_j)$, for $j \in [B]$.
%\eqref{proof_theorem2_pk} and \eqref{proof_theorem2_pk2},

Combining \eqref{unifrate4}, \eqref{lemma1} and \eqref{roptznm}, for given $v_{[B]}$, we obtain
\begin{align}
&R_{\rm{genie\_opt}}(K_{[B]}, n_{[B]}, M)\nonumber\\
&~~\geq \prod\limits_{j=1}^B\mathrm{Pr}\left(\mathrm{V}_j\geq v_j\right) \operatorname*{max}\limits_{\substack{z_j\in [\lceil \min\{f(n_j, v_j), v_j\}\rceil]\\j \in [B]}}\left\{\prod\limits_{j=1}^B\mathrm{Pr}\left(\mathrm{Z}_j\geq z_j\right)\sum\limits_{j=1}^B z_j \left(1-\frac{MB}{\min\limits_{j\in [B]}\lfloor\frac{n_j}{z_j}\rfloor}\right)\right\}.
\end{align}
For any $\widetilde{z}_{j}\in (0, f(n_j, v_j)]$ and $j\in [B]$, we have 
\begin{subequations}
\begin{align}
&R_{\rm{genie\_opt}}(K_{[B]}, n_{[B]}, M)\nonumber\\
&~~\geq \prod\limits_{j=1}^B\mathrm{Pr}\left(\mathrm{V}_j\geq v_j\right) \operatorname*{max}\limits_{\substack{z_j\in [\lceil \min\{\widetilde{z}_{j}, v_j\}\rceil]\\j \in [B]}}\left\{\prod\limits_{j=1}^B\mathrm{Pr}\left(\mathrm{Z}_j\geq z_j\right)\sum\limits_{j=1}^B z_j \left(1-\frac{MB}{\min\limits_{j\in [B]}\lfloor\frac{n_j}{z_j}\rfloor}\right)\right\}\label{ropt12}\\
&~~\geq \prod\limits_{j=1}^B\mathrm{Pr}\left(\mathrm{V}_j\geq v_j\right)\prod\limits_{j=1}^B\mathrm{Pr}\left(\mathrm{Z}_j\geq \widetilde{z}_{j}\right) \operatorname*{max}\limits_{\substack{z_j\in [\lceil \min\{\widetilde{z}_{j}, v_j\}\rceil]\\j \in [B]}}\left\{\sum\limits_{j=1}^B z_j \left(1-\frac{MB}{\min\limits_{j\in [B]}\lfloor\frac{n_j}{z_j}\rfloor}\right)\right\}\label{ropt13},
\end{align}
\end{subequations}
where \eqref{ropt12} is derived since $\widetilde{z}_{j}\leq f(n_j, v_j)$, $\forall j\in [B]$; \eqref{ropt13} follows by the fact that $z_j\leq \lceil\widetilde{z}_{j}\rceil$ and $Z_j$ is an integer, $\forall j \in [B]$. Using the lower bounds in \eqref{proof_theorem2_pk2} and \eqref{proof_theorem2_pz2}, and optimizing over $n_{[B]}$, $v_{[B]}$, and $\widetilde{z}_{[B]}$, we have 
\begin{align}\label{roptfi}
R_{\rm{opt}}(K_{[B]}, M)\geq & \operatorname*{max}\nolimits_{n_{[B]},~v_{[B]}, ~\widetilde{z}_{[B]}}\left\{\left(\prod\nolimits_{j=1}^Bf'_j(K_j, n_j, v_j)\right)\left(\prod\nolimits_{j=1}^Bf''_j(n_j, v_j, \widetilde{z}_j)\right)\cdot\right.\nonumber\\
&~~ \qquad\left.\operatorname*{max}\nolimits_{\substack{z_j\in [\lceil \min\{\widetilde{z}_j, v_j\}\rceil],j \in [B]}}\left\{\sum\nolimits_{j=1}^B z_j \left(1-\frac{MB}{\min\nolimits_{j\in [B]}\lfloor n_j/z_j\rfloor}\right)\right\}\right\},
\end{align}
where $n_j \in [N]$, $v_j\in (0, K_jn_jr_{n_jj}]$, $\widetilde{z}_{j}\in (0, f(n_j, v_j)]$, $j\in [B]$. Taking the expectation over all possible combinations of $(K_{[B]})$, we have
\begin{equation}
\begin{aligned}
R^* (M)\geq\mathbb{E}[R_{\rm{opt}}(K_{[B]}, M)]=\sum\nolimits_{k_{[B]}\in \mathcal{A}^B}\left(\prod\nolimits_{j=1}^B \mathrm{Pr}\{K_j=k_j\}\right)R_{\rm{opt}}(k_{[B]}, M),
\end{aligned}
\end{equation}
which, with \eqref{roptfi}, completes the proof of Theorem~\ref{lowerbound}.

\bibliographystyle{IEEEtran}
\bibliography{main}

\begin{thebibliography}{10}
\providecommand{\url}[1]{#1}
\csname url@samestyle\endcsname
\providecommand{\newblock}{\relax}
\providecommand{\bibinfo}[2]{#2}
\providecommand{\BIBentrySTDinterwordspacing}{\spaceskip=0pt\relax}
\providecommand{\BIBentryALTinterwordstretchfactor}{4}
\providecommand{\BIBentryALTinterwordspacing}{\spaceskip=\fontdimen2\font plus
\BIBentryALTinterwordstretchfactor\fontdimen3\font minus
  \fontdimen4\font\relax}
\providecommand{\BIBforeignlanguage}[2]{{%
\expandafter\ifx\csname l@#1\endcsname\relax
\typeout{** WARNING: IEEEtran.bst: No hyphenation pattern has been}%
\typeout{** loaded for the language `#1'. Using the pattern for}%
\typeout{** the default language instead.}%
\else
\language=\csname l@#1\endcsname
\fi
#2}}
\providecommand{\BIBdecl}{\relax}
\BIBdecl

\bibitem{yangaudience}
Q.~Yang, M.~{Mohammadi Amiri}, and D.~G\"und\"uz, ``Audience retention rate
  aware coded video caching,'' in \emph{Proc. {IEEE} Int'l Conf. Commun.
  Workshop (ICC Workshop)}, Paris, France, May 2017, pp. 1189--1194.

\bibitem{zeni2013youstatanalyzer}
M.~Zeni, D.~Miorandi, and F.~De~Pellegrini, ``Youstatanalyzer: A tool for
  analysing the dynamics of youtube content popularity,'' in \emph{Proc. of
  {ICST} VALUETOOLS}, Torino, Italy, Dec. 2013, pp. 286--289.

\bibitem{GolrezaeiFemtocaching}
N.~Golrezaei, A.~F. Molisch, A.~G. Dimakis, and G.~Caire, ``Femtocaching and
  device-to-device collaboration: A new architecture for wireless video
  distribution,'' \emph{{IEEE} Commun. Mag.}, vol.~51, no.~4, pp. 142--149,
  Apr. 2013.

\bibitem{GregoryDtoD}
M.~Gregori, J.~Gomez-Vilardebo, J.~Matamoros, and D.~G\"und\"uz, ``Wireless
  content caching for small cell and {D}2{D} networks,'' \emph{{IEEE} J. Sel.
  Areas Commun.}, vol.~34, no.~5, pp. 1222--1234, Mar 2016.

\bibitem{MaddahAliCentralized}
M.~A. Maddah-Ali and U.~Niesen, ``Fundamental limits of caching,'' \emph{{IEEE}
  Trans. Inform. Theory}, vol.~60, no.~5, pp. 2856--2867, May 2014.

\bibitem{samuel2017}
S.~O. Somuyiwa, A.~Gy\"orgy, and D.~G\"und\"uz, ``A reinforcement-learning
  approach to proactive caching in wireless networks,'' \emph{{IEEE} J. Sel.
  Areas Commun.}, to appear.

\bibitem{MaddahAliDecentralized}
M.~A. Maddah-Ali and U.~Niesen, ``Decentralized caching attains order optimal
  memory-rate tradeoff,'' \emph{{IEEE/ACM} Trans. Netw}, vol.~23, no.~4, pp.
  1029--1040, Apr. 2014.

\bibitem{MohammadQianDenizITW}
M.~{Mohammadi Amiri}, Q.~Yang, and D.~G\"und\"uz, ``Coded caching for a large
  number of users,'' in \emph{Proc. IEEE Inform. Theory Workshop (ITW)},
  Cambridge, UK, Sep. 2016.

\bibitem{MohammadDenizTCom}
M.~{Mohammadi Amiri} and D.~G\"und\"uz, ``Fundamental limits of coded caching:
  Improved delivery rate-cache capacity trade-off,'' \emph{{IEEE} Trans.
  Commun.}, vol.~65, no.~2, pp. 806--815, Feb. 2017.

\bibitem{JiArXivNonuniform}
M.~Ji, A.~M. Tulino, J.~Llorca, and G.~Caire, ``Order-optimal rate of caching
  and coded multicasting with random demands,'' \emph{{IEEE} Trans. Inform.
  Theory}, vol.~63, no.~6, pp. 3923--3949, Jun. 2017.

\bibitem{NiesenNonuniform}
U.~Niesen and M.~A. {Maddah-Ali}, ``Coded caching with nonuniform demands,''
  \emph{{IEEE} Trans. Inform. Theory}, vol.~63, no.~2, pp. 1146--1158, Feb.
  2017.

\bibitem{PedarsaniOnlineCaching}
R.~Pedarsani, M.~A. Maddah-Ali, and U.~Niesen, ``Online coded caching,'' in
  \emph{Proc. {IEEE} Int'l Conf. Commun. (ICC)}, Sydney, Australia, Jun. 2014,
  pp. 1878--1883.

\bibitem{yang2016coded}
Q.~Yang and D.~G\"und\"uz, ``Coded caching and content delivery with
  heterogeneous distortion requirements,'' \emph{{IEEE} Trans. Inform. Theory},
  vol.~64, no.~6, pp. 4347--4364, Jun. 2018.

\bibitem{DistinctAmiriYangGunduz}
M.~{Mohammadi Amiri}, Q.~Yang, and D.~G\"und\"uz, ``Decentralized caching and
  coded delivery with distinct cache capacities,'' \emph{{IEEE} Trans.
  Commun.}, vol.~65, no.~11, pp. 4657--4669, Nov. 2017.

\bibitem{maggi2015adapting}
L.~Maggi, L.~Gkatzikis, G.~Paschos, and J.~Leguay, ``Adapting caching to
  audience retention rate: Which video chunk to store?'' \emph{{Comput.
  Commun.}}, vol. 116, pp. 159--171, Jan. 2018.

\bibitem{emreicc2018}
E.~Ozfatura and D.~G\"und\"uz, ``Uncoded caching and cross-level coded delivery
  for non-uniform file popularity,'' in \emph{Proc. {IEEE} Int'l Conf. Commun.
  (ICC)}, Kansas City, MO, May 2018.

\bibitem{wang2015optimal}
L.~Wang, S.~Bayhan, and J.~Kangasharju, ``Optimal chunking and partial caching
  in information-centric networks,'' \emph{{Comput. Commun.}}, vol.~61, pp.
  48--57, May 2015.

\bibitem{breslau1999web}
L.~Breslau, P.~Cao, L.~Fan, G.~Phillips, and S.~Shenker, ``Web caching and
  {Zipf}-like distributions: Evidence and implications,'' in \emph{Proc. {IEEE}
  Conf. Comput. Commun. (INFOCOM)}, NY, Mar. 1999, pp. 126--134.

\bibitem{yu2006dynamic}
J.~Yu, C.~T. Chou, Z.~Yang, X.~Du, and T.~Wang, ``A dynamic caching algorithm
  based on internal popularity distribution of streaming media,''
  \emph{Multimedia Syst.}, vol.~12, no.~2, pp. 135--149, Jul. 2006.

\bibitem{CoverInformation}
T.~M. Cover and J.~A. Thomas, \emph{Elements of Information Theory}.\hskip 1em
  plus 0.5em minus 0.4em\relax New York, NY, USA: Wiley, 1991.

\end{thebibliography}

\end{document}